\documentclass[12pt]{article}
\newcommand{\beq}{\begin{equation}}
\newcommand{\eeq}{\end{equation}}
\newcommand{\beqa}{\begin{eqnarray}}
\newcommand{\eeqa}{\end{eqnarray}}
\newcommand{\beqar}{\begin{eqnarray*}}
\newcommand{\eeqar}{\end{eqnarray*}}

\newcommand{\eps}{\epsilon}
\newcommand{\ga}{\gamma}
\newcommand{\Ga}{\Gamma}

\newcommand{\inn}{\!\cdot\!}

\newcommand{\z}{\zeta}

\newcommand{\eg}{{\it e.g.,}\ }
\newcommand{\ie}{{\it i.e.,}\ }
\newcommand{\labell}[1]{\label{#1}} %{\label{#1}} %
\newcommand{\reef}[1]{(\ref{#1})}
\newcommand\prt{\partial}
\newcommand\veps{\varepsilon}

\newcommand\cF{{\cal F}}

\newcommand\cA{{\cal A}}

\newcommand\cM{{\cal M}}

\newcommand\cP{{\cal P}}

\newcommand\cB{{\cal B}}

\newcommand\bz{\bar{z}}

\newcommand\Tr{{\rm Tr}}

\begin{document}

\vspace*{1cm}

\begin{center}
{\bf \Large   Tree-level  S-matrix elements from  S-duality }

\vspace*{1cm}

{Mohammad R. Garousi\footnote{garousi@ferdowsi.um.ac.ir} }\\
\vspace*{1cm}
{ Department of Physics, Ferdowsi University of Mashhad,\\ P.O. Box 1436, Mashhad, Iran}
\\
\vspace{2cm}

\end{center}

\begin{abstract}
\baselineskip=18pt

It has been  speculated  that  the S-matrix elements of type IIB superstring theory  satisfy the Ward identity associated with the S-duality. This indicates that a group of S-matrix elements at each loop level are invariant  under the linear $SL(2,R)$ transformations. If one evaluates one component of such S-dual multiplet, then all other components can be found  by the simple use of  the linear $SL(2,R)$ transformations. 

In this paper, we calculate the disk-level S-matrix element of one graviton(dilaton), one B-field and one gauge boson  on the word volume of D$_3$-brane. The S-dual multiplet corresponding to the graviton(dilaton) amplitude has three(six) components. In particular, the graviton multiplet  has the S-matrix element of  one graviton, one R-R two-form and one gauge boson, and the dilaton multiplet  has the S-matrix element of  one R-R scalar, one B-field and one gauge boson vertex operators. We calculate explicitly these particular components  and show that they are  precisely  the ones  predicted  by  the S-duality. We have also found  the low energy contact terms of the dilaton multiplet at order $\alpha'^2$.

\end{abstract}
Keywords: S-duality, S-matrix, D$_3$-brane

\vfill
\setcounter{page}{0}
\setcounter{footnote}{0}
\newpage

%\section{The idea} \label{intro}

\section{Introduction } \label{intro}
It is known that the  type IIB superstring theory  is invariant  under S-duality \cite{Font:1990gx,Sen:1994fa,Rey:1989xj,Sen:1994yi,Schwarz:1993cr,Hull:1994ys,Becker:2007zj}. This symmetry  should be carried by  the S-matrix elements.  It has been speculated in \cite{Garousi:2011we} that the linear S-duality should appear in the tree-level S-matrix elements through the associated Ward identity. This classifies the tree-level S-matrix elements into S-dual multiplets. Each multiplet includes S-matrix elements which interchange under the linear $SL(2,R)$ transformation. The S-matrix elements of gravitons are singlets in this classification.   The  tree-level S-dual multiplets should then be dressed with the loops and the nonperturbative effects to becomes  invariant under the full nonlinear S-duality transformations \cite{Green:1997tv} - \cite{Garousi:2011fc}.

The S-duality holds order by order in $\alpha'$ and is nonperturbative in the string loop expansion \cite{Becker:2007zj}.  Therefore,  one has to  $\alpha'$-expand an amplitude in the Einstein frame and then study its S-duality at each order of $\alpha'$. Let us consider the  sphere-level S-matrix element of four gravitons  whose S-duality  has been studied in \cite{Green:1997tv} - \cite{Basu:2007ck}. The leading $\alpha'$-order contact term of this amplitude which has eight momenta is not     
 invariant under the S-duality because of the presence of the dilaton factor $e^{-3\phi_0/2}$ in these couplings in the Einstein frame. It has been conjectured that the dilaton  factor might be extended to the non-holomorphic Eisenstein series $E_{3/2}(\phi_0,C_0)$ after including the loops and the nonperturbative effects  \cite{Green:1997tv}. Similar conjecture has been made for the  higher $\alpha'$ order terms  \cite{Green:1997tv} - \cite{Basu:2007ck}. However, the leading  term of this expansion which is a massless pole,  has no dilaton factor so this term by itself  is invariant under the S-duality. 
 
 The above proposal for constructing the S-dual S-matrix element of four gravitons  has been extended to the disk-level S-matrix element of two gravitons  in \cite{ Bachas:1999um,Basu:2008gt},  to the disk-level S-matrix elements of some other external states in \cite{Garousi:2011fc,Garousi:2011we,Garousi:2011vs} and to the nonabelian S-matrix elements   on the world volume of multiple D$_3$-branes in \cite{Garousi:2011jh}. The leading  $\alpha'$ order term in each of the abelian S-matrix elements    is invariant under the S-duality without dressing it with the loops and the nonperturbative effects. We will show that such property is hold by the S-matrix elements that we will find in this paper as well.
 
 In general, the evaluation of the tree-level S-matrix element of the Ramond-Ramond (R-R) vertex operators  is much more complicated than the evaluation of the tree-level S-matrix element of the Neveu Schwarz-Neveu Schwarz (NS-NS) vertex operators because of the presence of the spin operator in the R-R vertex operator \cite{Cohn:1986bn}
. The S-duality relates the R-R two-form to the B-field and the R-R scalar to the dilaton. Using the observation that the S-matrix elements should satisfy the Ward
 identity corresponding to the S-duality \cite{Garousi:2011we}, one can relate the S-matrix element of the R-R vertex operators to the S-matrix element of the NS-NS vertex operators. In this paper we would like to examine this idea by evaluating the disk-level S-matrix element of one graviton(dilaton), one B-field and one gauge boson and relating it to various S-matrix elements involving the R-R vertex operators.  
 
 The outline of the paper is as follows: In section 2, using the conformal field theory technique, we calculate explicitly the disk-level scattering amplitude of one graviton, one B-field and one gauge boson. In section 2.1, we find the S-dual multiplet corresponding to this amplitude which has three components. We will show that the leading $\alpha'$-order terms of the multiplet are invariant under the S-duality. In section 3, we calculate the dilaton amplitudes corresponding to the above multiplet by replacing the graviton polarization with the dilaton polarization tensor. The dilaton amplitudes by themselves, however, do not form an S-dual multiplet because unlike the graviton the dilaton is not invariant under the S-duality. In section 3.1, we find the S-dual multiplet corresponding to the dilaton amplitudes. In section 4, we examine some of the amplitudes that we have found from the S-duality with explicit calculation and find exact agreement. In section 5, we find the low energy contact terms of the dilaton multiplet at order $\alpha'^2$.
 
 \section{Graviton amplitude}
 
The tree-level scattering amplitude of two gravitons and one transverse scalar has been calculated in \cite{Fotopoulos:2001pt}. In this section we are interested in the scattering amplitude of   one graviton, one B-field and one gauge boson   on the world-volume of a D$_p$-brane. This amplitude   is given  by the correlation function of their  corresponding vertex operators on  disk. Since the background charge of the world-sheet with topology of a disk is $Q_{\phi}=2$ one has to choose  the vertex operators in the appropriate pictures to produce the compensating charge $Q_{\phi}=-2$.
 The scattering amplitude may then be given  by the following correlation function:
\beqa
A&\sim&<V_{G}^{(0,0)}(\veps_1,p_1)V_{B}^{(-1,-1)}(\veps_2,p_2)V_{A}^{(0)}(\z_3,k_3)>\labell{amp21}
\eeqa
Using the doubling trick \cite{Garousi:1996ad,Hashimoto:1996bf}, the vertex operators are given by the following integrals on the upper half $z$-plane:\footnote{Our conventions set $\alpha'=2$. Our index convention is that the Greek letters  $(\mu,\nu,\cdots)$ are  the indices of the space-time coordinates, the Latin letters $(a,d,c,\cdots)$ are the world-volume indices and the letters $(i,j,k,\cdots)$ are the normal bundle indices.}
\beqa
V_{G}^{(0,0)}&\!\!\!\!\!=\!\!\!\!\!&(\veps_1\inn D)_{\mu_1\nu_1}\int d^2z_1:(\prt X^{\mu_1}+ip_1\inn\psi\psi^{\mu_1})e^{ip_1\cdot X}:(\prt X^{\nu_1}+ip_1\inn D\inn\psi\psi^{\nu_1})e^{ip_1\cdot D\cdot X}:\nonumber\\
V_{B}^{(-1,-1)}&\!\!\!\!\!=\!\!\!\!\!&(\veps_2\inn D)_{\mu_2\nu_2}\int d^2z_2:\psi^{\mu_2}e^{-\phi}e^{ip_2\cdot X}:\psi^{\nu_2}e^{-\phi}e^{ip_2\cdot D\cdot X}:\nonumber\\
V_{A}^{(0)}&\!\!\!\!\!=\!\!\!\!\!&(\z_3)_{a}\int dx_3:(\prt X^{a}+2ik_3\inn\psi\psi^{a})e^{2ik_3\cdot X}
\eeqa
where  the matrix $D^{\mu}_{\nu}$ is diagonal with $+1$ in the world volume directions and $-1$ in the transverse direction. The polarization $\veps_1$ is symmetric and the polarization $\veps_2$ is antisymmetric.  We will not impose the traceless condition for the graviton polarization. This will allow us in the next section to find the dilaton amplitude from the graviton amplitude  by choosing the graviton  polarization to be the flat metric $\eta_{\mu\nu}$. The on-shell conditions are $\veps_i.p_i=p_i.\veps_i=p_i.p_i=0$ for $i=1,2$, and $\z_3.k_3=k_3.k_3=0$.

Using the standard world-sheet propagators
\beqa
<X^{\mu}(x)X^{\nu}(y)>&=&-\eta^{\mu\nu}\log(x-y)\nonumber\\
<\psi^{\mu}(x)\psi^{\nu}(y)>&=&-\frac{\eta^{\mu\nu}}{x-y}\nonumber\\
<\phi(x)\phi(y)>&=&-\log(x-y)\labell{wpro}
\eeqa
one can calculate the correlators in \reef{amp21}. The result is an expression which satisfies the Ward identity associated with the gauge boson, \ie if one replaces the gauge boson polarization $\z_3$ with its momentum $k_3$ the result is zero. So one can write the result in terms of the gauge boson field strength $F^{ab}=i(k_3^a\z_3^b-k_3^b\z_3^a)$. The result in the string frame  is 
\beqa
A&\!\!\!\!\!\sim\!\!\!\!\!&2T_p\bigg[2 p_1.F.\veps_1.\veps_2.p_1 I_1+\left(-2 p_1.F.\veps_2.\veps_1.k_3+p_1.F.\veps_1.D.\veps_2.p_1+p_1.F.\veps_1.\veps_2.D.p_1\right) I_3\nonumber\\&&+\left(2 p_1.\veps_2.F.\veps_1.k_3+p_1.\veps_2.F.\veps_1.D.p_1\right) I_3+p_1.F.\veps_2.\veps_1.p_2 I_4-p_1.\veps_2.F.\veps_1.p_2 I_4\nonumber\\&&-p_1.F.\veps_2.\veps_1.D.p_1 I_5-2 p_1.D.\veps_1.k_3 I_6 \Tr\left[F.\veps_2\right]+p_2.\veps_1.p_2 I_8 \Tr\left[F.\veps_2\right]-\nonumber\\&&2 I_2 \left(2 k_3.\veps_1.F.\veps_2.D.p_1-2 p_1.F.\veps_2.D.\veps_1.k_3+p_1.D.\veps_1.F.\veps_2.D.p_1+p_1.F.\veps_1.D.\veps_2.D.p_1\right.\nonumber\\&&\left.+2 k_3.\veps_1.k_3 \Tr\left[F.\veps_2\right]\right)+I_{10} \left(p_1.F.\veps_2.D.\veps_1.p_2-p_2.\veps_1.F.\veps_2.D.p_1-2 p_2.\veps_1.k_3 \Tr\left[F.\veps_2\right]\right)\nonumber\\&&+I_{11} \left(-2 p_1.F.\veps_2.D.\veps_1.D.p_1+\left(p_1.D.\veps_1.p_2+p_2.\veps_1.D.p_1+p_1.D.\veps_1.D.p_1\right) \Tr\left[F.\veps_2\right]\right)\nonumber\\&&+\left(-\left(p_1.p_2+p_1.D.p_1\right) \left(I_2-I_6\right)+p_1.k_3 I_6\right) \Tr\left[F.\veps_2\right] \Tr\left[\veps_1.D\right]-I_7 \left(p_1.F.\veps_2.p_1 \Tr\left[\veps_1.D\right]\right.\nonumber\\&&\left.+p_1.D.p_1 \Tr\left[F.\veps_1.\veps_2\right]\right)+I_9 \left(p_1.F.\veps_2.D.p_1 \Tr\left[\veps_1.D\right]+p_1.D.p_1 \Tr\left[F.\veps_1.D.\veps_2\right]\right)\bigg]\labell{amp20}
\eeqa
where  $I_1,\cdots I_{11}$ are some integrals. The amplitude has also a delta Dirac function imposes the momentum conservation along the brane, \ie
\beqa
p_1+p_1.D+p_2+p_2.D+2k_3&=&0
\eeqa
 To have consistency with the T-duality, the amplitude should have no extra dilaton factor. The dilaton appears only in the brane tension which is consistent with T-duality, \ie $T_p\delta^{p+1}(\cdots)\rightarrow T_{p-1}\delta^p(\cdots)$. We will not fix the overall numeric factor of the amplitude. The explicit form of the integrals $I_2,\, I_{11}$ are the following:
\beqa
I_2&=&\frac{K}{z_{31} z_{2 \bar{1}} z_{3 \bar{1}} z_{1\bar{2}} z_{2 \bar{2}}}\nonumber\\
I_{11}&=&\frac{K}{z_{32}z_{1\bar{1}} z_{2 \bar{1}} z_{1\bar{2}} z_{3 \bar{2}}}\labell{int.1}
\eeqa
and the other integrals appear in the Appendix. The notation $z_{ij}$ is $z_{ij}=z_i-z_j$. There is a measure $\int d^2z_1d^2z_2dx_3$ for all the integrals which we have omitted.  The function $K$ is
\beqa
K&=&z_{ 1\bar{1}}^{p_1.D.p_1}|z_{12}|^{2p_1.p_2}|z_{ 1\bar{2}}|^{2p_1.D.p_2}|z_{13}|^{4p_1.k_3}z_{ 2\bar{2}}^{p_2.D.p_2}|z_{23}|^{4p_2.k_3}
\eeqa
Note that the integral $I_{11}$ is the same as $I_2$ in which the labels $1,2$ are interchanged.

One may use the relation between disk amplitudes involving mixed open and closed strings, and disk amplitudes with only open strings \cite{Garousi:1996ad,Hashimoto:1996bf,Stieberger:2009hq} to relate the  integrals \reef{int.1} to the integrals that appear in the five open string amplitude \cite{Kitazawa:1987xj}. However, we prefer to calculate the integrals \reef{int.1} directly which is easy to perform. The integrals are invariant under the $SL(2,R)$ transformation of the upper half-plane. Using this symmetry one can map the result to unit disk, and then fix the symmetry by fixing $z_1=0$ and $x_3=1$. In the polar coordinate $z_2=re^{i\theta}$, the $\theta$-integral can be evaluated using the following formula \cite{ISG}:
\beqa
\int_0^{2\pi}d\theta\frac{\cos(n\theta)}{(1+x^2-2x\cos(\theta))^b}&=&2\pi x^n\frac{\Ga(b+n)}{n!\Ga(b)}{}_2F_1\bigg[{b, \ n+b \atop n+1}\ ;\ x^2\bigg]
\eeqa
where $|x|<1$, and the $r$-integral can be evaluated using the following formula \cite{ISG}:
\beqa
{}_{3}F_{2}\bigg[{1+a,\ a_1, \  a_2\atop  2+a+b,\ b_1}\ ;\ 1
\bigg]B(1+a,1+b)&\!\!\!\!\!=\!\!\!\!\!&\int_0^1dx\,x^a(1-x)^b{}_{2}F_{1}\bigg[{ a_1, \  a_2\atop   b_1}\ ;\  x
\bigg]\nonumber
\eeqa
The integral  $I_2$ then becomes
\beqa
I_2(p_1,p_2,k_3)&=&\pi B(1+p_1.p_2,p_2.D.p_2)\,{}_{3}F_{2}\bigg[{1+p_1.p_2,\ 2p_1.k_3, \  2p_1.k_3\atop 1+ p_1.p_2+p_2.D.p_2,\ 1}\ ;\ 1
\bigg]\labell{int.11}
\eeqa
%Note that the structure of this integral is the same as the integral that appears in the disk amplitude of five open strings \cite{...}. 
The integral $I_{11}$ is
\beqa
I_{11}(p_1,p_2,k_3)&=&I_2(p_2,p_1,k_3)
\eeqa
 There are four Mandelstam variables $p_1.p_2,\,p_2.D.p_2, \, p_1.D.p_1$ and $p_1.k_3$ in $I_2$ and $I_{11}$. There are, however, three  physical channels. $p_1.p_2$-channel which is a closed string channel, and $p_2.D.p_2$-channel and $p_1.D.p_1$-channel which are open string channels. The $p_1.D.p_1$-channel in $I_2$ is coming from the Hyper-geometric function.  There is no open or closed string channel corresponding to the Mandelstam variable $p_1.k_3$. The integral $I_2$ has massless pole only in the $p_2.D.p_2$-channel.
To study the low energy limit of the scattering amplitudes in section 5, we need the  $\alpha'$-expansion of  the integrals which are given by 
\beqa
I_2&=&\frac{1}{p_2.D.p_2 }+\frac{1}{6} \pi ^2 \left(-p_1.p_2+\frac{4 (p_1.k_3)^2}{p_2.D.p_2}\right) +\cdots\labell{expand}\\
I_{11}&=&\frac{1}{p_1.D.p_1 }+\frac{1}{6} \pi ^2 \left(-p_1.p_2+\frac{4 (p_1.k_3)^2}{p_1.D.p_1}\right) +\cdots\nonumber
\eeqa
where we have also used the relation $p_2.k_3=-p_1.k_3$. We have used the package \cite{Huber:2005yg} for expanding the Hypergeometric function ${}_3F_2$.

If one calculate the other integrals, one would  find similar results for them. Then one may use these functions to find some relations between  them. Using these relations, one would find that the amplitude would satisfy the Ward identity associated with the graviton and B-field. Alternatively, one may impose the graviton and the B-field Ward identities to find these relations. We follow this latter method to find the  relations:
\beqa
&&I_{10}=2 I_2+I_3,\,\,I_4=2 I_1-I_3,\,\,I_8=I_1-I_2-I_3,\nonumber\\
&&I_7=I_2+I_{11}+I_3,\,\,I_5=I_3-2I_7,\,\,I_9=I_2+I_{11},I_6=I_2-I_{11}\labell{rel.1}
\eeqa
which we have checked with the explicit form of the integrals. One also finds the following two equations involving the Mandelstam variables:
\beqa
&&4 I_2 k_3.p_1+\left(2 I_2+I_3\right) p_1.p_2+\left(I_2-I_{11}\right) p_1.D.p_1=0,\nonumber\\
&&2 I_3 k_3.p_1+\left(-2 I_1+I_3\right) p_1.p_2-\left(I_2+I_{11}\right) p_1.D.p_1=0\labell{rel.2}
\eeqa
which may be verified using by part integration. Using the above equations \reef{rel.1} and \reef{rel.2}, one can write $I_1,I_3,\cdots I_{10}$ in terms of $I_2$ and $I_{11}$. The result is a lengthy expression. To compress the result, one may   write the amplitude \reef{amp20}    in terms of  field strength of the B-field, \ie $H=dB$. Note that there is no unique way to write the amplitude in terms of $H$, \ie using the on-shell condition one can find different expressions for the amplitude in term of $H$.  We impose the condition that the final amplitude at low energy should have no double  pole  $1/(p_1.p_2p_1.D.p_1)$ which is consistent with the low energy effective action. Using this condition, we find  
  \beqa
A_{gBF}&\sim&T_p(I_{11}A_{11}+I_2A_2)\labell{ampg}
\eeqa
where $A_{11}$ is
\beqa
A_{11}&=& F^{ab} \bigg[-4 p_1.D.\veps_1.N.H_{ab}+2p_1.N.H_{ab} \Tr\left[\veps_1.D\right]+\frac{ p_1.D.p_1}{p_1.p_2}\left(\frac{}{}2 k_3.\veps_1.H_{ab}\right.\nonumber\\
&&\left.-4 p_1.H_{\mu a} \left(\veps_1.N\right)_{b}{}^{\mu}+2p_2.\veps_1.N.H_{ab}-p_1.D.\veps_1.H_{ab}+p_1.H_{ab} \Tr\left[\veps_1.D\right]\frac{}{}\right)\nonumber\\
&&-\frac{ 2p_1.D.p_1}{(p_1.p_2)^2}\left(\frac{}{}p_1.k_3 \,p_2.\veps_1.H_{ab} -p_1.D.H_{\mu a} (p_2.\veps_1)_b \left(p_1\right)^{\mu }\right.\nonumber\\&&\left.+p_1.D.p_1\,p_1.H_{\mu a}  \left(\veps_1\right)_{b}{}^{\mu}+p_1.D.H_{\mu \nu } (p_1)_a\left(\veps_1\right)_{b}{}^{\mu } \left(p_1\right)^{\nu }\frac{}{}\right)\bigg]\labell{ampg.1}
\eeqa
 and $A_2$ is 
 \beqa
A_2&\!\!\!\!\!=\!\!\!\!\!&F^{ab}\bigg[-8 k_3.\veps_1.V.H_{ab}-2p_1.V.H_{ab} \Tr\left[\veps_1.D\right] -16 p_1.V.H_{ca}(\veps_1)_b{}^c -\frac{p_2.D.p_2}{p_1.p_2} \left(\!\!\frac{}{} 2 k_3.\veps_1.H_{ab} \right.\nonumber\\
&&\left.+4 p_1.H_{\mu a} \left(\veps_1.V\right)_{b}{}^{\mu}-2p_2.\veps_1.V.H_{ab} -p_1.D.\veps_1.H_{ab}+p_1.H_{ab} \Tr\left[\veps_1.D\right] \frac{}{}\right)\nonumber\\
&&+\frac{2 p_2.D.p_2}{(p_1.p_2)^2}\left(\frac{}{}p_1.k_3 \,p_2.\veps_1.H_{ab} -p_1.D.H_{\mu a} (p_2.\veps_1)_b \left(p_1\right)^{\mu }\right.\nonumber\\&&\left.+p_1.D.p_1\,p_1.H_{\mu a}  \left(\veps_1\right)_{b}{}^{\mu}+p_1.D.H_{\mu \nu } (p_1)_a\left(\veps_1\right)_{b}{}^{\mu } \left(p_1\right)^{\nu }\frac{}{}\right)\bigg]\labell{ampg.2}
\eeqa
where $H^{\mu\nu\rho}=i(p_2^{\mu}\veps_2^{\nu\rho}+p_2^{\rho}\veps_2^{\mu\nu}+p_2^{\nu}\veps_2^{\rho\mu})$, the matrix $V$ is the flat metric of the world volume, \ie $V^{ab}=\eta^{ab}$, and $N$ is the flat metric of the transverse space, \ie $N^{ij}=\eta^{ij}$.  

The amplitude does not satisfy the Ward identity corresponding to the graviton unless one rewrite the field strength $H$ in terms of $B$. So one can not write the amplitude in terms of  field strengths $F$, $H$ and the curvature $R$. However, some of them can be written as $RHF$. For example, the terms that have $p_1.D.H_{\mu\nu}$ satisfy the Ward identity associated with the graviton, so they can be rewritten as $RHF$. One should not expect that all terms to be rewritten as $RHF$ because the metric can appear as contracting the indices of the covariant  derivatives of $\sqrt{-g}HF$ as well as in the definition of the covariant derivatives. In general, using the fact that the contact terms of the S-matrix elements must produce couplings in effective field theory which must be  in terms of covariant derivative of field strengths,  one expects a  S-matrix element to be written in terms of field strengths except for the graviton. For graviton, only some terms of the amplitude can be written as curvature. 

As a check of the result in \reef{ampg}, we note that if one examines the Feynman diagrams in low energy field theory, there would be no double pole $1/(p_1.p_2)^2$. Therefore, the double pole in the amplitude \reef{ampg} must be canceled. Using the expansion \reef{expand}, one observes that the double pole in $A_{11}$ and $A_2$ in fact is canceled in the whole amplitude \reef{ampg}. We have expand the integrals $I_2$ and $I_{11}$ up to $(\alpha')^6$ and found no double pole. Moreover, taking in to account that the integral $I_2$ has simple pole in $p_2.D.p_2$ and the integral $I_{11}$ has simple pole in $p_1.D.p_1$, one observes that the amplitude has no double pole  $1/(p_1.p_2p_1.D.p_1)$ or $1/(p_1.p_2p_2.D.p_2)$ either. 

%the expressions $A_1$ and $A_2$ simplifies to
%\beqa
%A_2&=&\frac{2(p_1.p_2+p_1.D.p_1+3p_1.D.p_2)}{p_1.k_3}p_1.F.\veps _1.\veps _2.p_1 \nonumber\\
%A_1&=&-\frac{2(p_1.D.p_2+p_1.D.p_1+3p_1.p_2)}{p_1.k_3}p_1.F.\veps _1.\veps _2.p_1 
%\eeqa
 
 \subsection{Graviton S-dual multiplet}
 
To study the S-duality of the amplitude \reef{ampg}, we have to convert it to the Einstein frame.  The relation between the string frame metric which is used in \reef{ampg}, and the Einstein frame is  $g^S_{\mu\nu}=e^{\phi_0/2}g^E_{\mu\nu}$. The amplitudes \reef{ampg} in the Einstein frame is 
 \beqa
 A_{gBF}&\sim&I_{11}T_3e^{-3\phi_0/2}F^{ab}\bigg[-4 p_1.D.\veps_1.N.H_{ab}+\cdots\bigg]\nonumber\\
 &&+
 I_2T_3e^{-3\phi_0/2}F^{ab}\bigg[-8 k_3.\veps_1.V.H_{ab}+\cdots\bigg]\labell{amp25}
 \eeqa
where dots represent the other terms in \reef{ampg.1} and \reef{ampg.2}. The dilaton factor $e^{-\phi_0/2}$ must be multiplied to each  Mandelstam variables in the integrals $I_{11},\, I_2$ in the Einstein frame. For example,  $p_1.p_2$ in the string frame must be replaced by $ p_1.p_2\,e^{-\phi_0/2}$ in the Einstein frame. 

Under  S-duality, graviton is invariant and the following objects transform as doublets \cite{Gibbons:1995ap,Tseytlin:1996it,Green:1996qg}:
\beqa
\cB&\equiv&\pmatrix{B \cr 
C^{(2)}}\rightarrow (\Lambda^{-1})^T \pmatrix{B \cr 
C^{(2)}}\,\,\,;\,\,\,\Lambda=\pmatrix{p&q \cr 
r&s}\labell{2}\\
\cF&\equiv&\pmatrix{(*F) \cr 
e^{-\phi_0}F-C_0(*F)}\rightarrow (\Lambda^{-1})^T \pmatrix{(*F) \cr 
e^{-\phi_0}F-C_0(*F)}\nonumber
\eeqa
where $(*F)_{ab}=\eps_{abcd}F^{cd}/2$, and $\phi_0$ and $C_0$ are the constant dilaton and R-R scalar, respectively. Consider  the  $SL(2,R)$ matrix $\cM$ 
\beqa
 {\cal M}=e^{\phi}\pmatrix{|\tau|^2&C \cr 
C&1}\labell{M}
\eeqa
where $C$ is the R-R scalar and $\tau=C+ie^{-\phi}$. This matrix  transforms under the $SL(2,R)$ transformation as\footnote{Note that the matrix $\cM$ here is the inverse of the matrix $\cM$ in \cite{Gibbons:1995ap}.}
\beqa
{\cal M}\rightarrow \Lambda {\cal M}\Lambda ^T
\eeqa
When the dilaton and the R-R scalar are constant, one finds  the following S-dual multiplet: 
\beqa
(*\cF)^T\cM_0\cB&=&-e^{-\phi_0}FB+(*F)C^{(2)}+C_0(*F)B\labell{inv.1}
\eeqa
where   the matrix $\cM_0$ is the matrix $\cM$ in which the dilaton and the R-R scalar are constant. The minus of amplitude \reef{amp25} corresponds to the first component of  the above multiplet. The  amplitude corresponding to the second component is 
\beqa
 A_{gC^{(2)}F}&\sim&\frac{1}{2}I_{11}T_3e^{-\phi_0/2}\eps^{abcd}F_{cd}\bigg[-4 p_1.D.\veps_1.N.F^{(3)}_{ab}+\cdots\bigg]\nonumber\\
 &&+\frac{1}{2}
 I_2T_3e^{-\phi_0/2}\eps^{abcd}F_{cd}\bigg[-8 k_3.\veps_1.V.F^{(3)}_{ab}+\cdots\bigg]\labell{amp26}
 \eeqa
where $F^{(3)}=dC^{(2)}$. Finally, the amplitude corresponding to the third component of the multiplet \reef{inv.1} is
\beqa
 A_{gBFC^{(0)}}&\sim&\frac{1}{2}I_{11}T_3e^{-\phi_0/2}\eps^{abcd}F_{cd}C_0\bigg[-4 p_1.D.\veps_1.N.H_{ab}+\cdots\bigg]\nonumber\\
 &&+\frac{1}{2}
 I_2T_3e^{-\phi_0/2}\eps^{abcd}F_{cd}C_0\bigg[-8 k_3.\veps_1.V.H_{ab}+\cdots\bigg]\labell{amp27}
 \eeqa
In above amplitude,  the R-R scalar must be constant. The  polarization of this constant  R-R scalar  is denoted by  $C_0$. This polarization is one, but for clarity we keep it as $C_0$. 

The amplitudes \reef{amp26} and \reef{amp27}  are the S-duality prediction for the S-matrix elements. The   amplitude \reef{amp27} should appear in the  S-matrix element of one R-R scalar, one graviton, one B-field and one gauge boson which is related to the amplitude of seven open strings \cite{Stieberger:2009hq}. In the limit that the momentum of the R-R scalar goes to zero, this amplitude should be reduced to \reef{amp27} and some massless poles. We leave the details of this calculation  for the future works. The amplitude \reef{amp26}, however,  is a 5-point function. In section 4 we explicitly calculate   the disk-level S-matrix element of one graviton, one R-R two-form and one gauge boson. We will show that the result is precisely agree with the amplitude \reef{amp26}.

Using the fact that the leading $\alpha'$ order terms of the integrals $I_2,\, I_{11}$ are massless pole, there is a dilaton factor $e^{\phi_0/2}$ in the leading order terms of these integrals in the Einstein frame. Therefore, the leading order terms of the amplitudes \reef{amp26}, \reef{amp27} have no dilaton factor and the leading order term of the amplitude \reef{amp25} has the dilaton factor $e^{-\phi_0}$. They are consistent with the multiplet \reef{inv.1}. This indicates that the leading $\alpha'$-order terms of the multiplet \reef{amp25}-\reef{amp26}-\reef{amp27} are invariant under the S-duality. It seems this symmetry is carried by any  abelian tree-level S-matrix elements.  The extra dilaton factors in all higher order terms may be dressed with the loops and the nonperturbative effects \cite{Green:1997tv} - \cite{Garousi:2011fc} to become invariant under the full nonlinear S-duality.

 \section{Dilaton amplitude}
 
 The dilaton amplitude can be read from the graviton amplitude by replacing the graviton polarization with
 \beqa
 (\veps_1)_{\mu\nu}&=&\eta_{\mu\nu}-\ell_{\mu}(p_1)_{\nu}-\ell_{\nu}(p_1)_{\mu}
 \eeqa
 where the auxiliary field $\ell$ satisfies $\ell.p_1=1$ and should be canceled in the final amplitude. The graviton amplitude \reef{amp20} satisfies the Ward identity, \ie if one replaces the graviton polarization $(\veps_1)_{\mu\nu}$ with $\z_{\mu}(p_1)_{\nu}+\z_{\nu}(p_1)_{\mu}$ where $\z_{\mu}$ is an arbitrary vector, the amplitude becomes zero. Since we have not used the traceless condition for the graviton,  it is obvious that the  replacement $-\ell_{\mu}(p_1)_{\nu}-\ell_{\nu}(p_1)_{\mu}$ for the graviton polarization gives zero result. %As we discussed before,  the amplitude \reef{ampg} does not satisfy the Ward identity corresponding to the graviton unless one rewrite the amplitude in terms of the B-field polarization instead of the field strength $H$. 
 So to find the dilaton amplitude we have to replace the graviton polarization in the amplitude \reef{amp20}  with $\eta_{\mu\nu}$. The amplitude can then be written in terms of $H$. The result for D$_3$-brane in the Einstein frame is
 \beqa
 A_{\phi BF}&\sim&4\phi_1T_3e^{-3\phi_0/2}F^{ab}\left(I_{11}\bigg[\frac{p_1.D.p_1}{p_1.p_2}  p_1.V.H_{ba}+\frac{p_1.k_3\,p_1.D.p_1}{(p_1.p_2)^2}p_1.H_{ba}\bigg]\right.\labell{ampphi}\\
 &&\left.-I_2\bigg[4p_1.V.H_{ba}+\frac{p_2.D.p_2}{p_1.p_2} \left(2 p_1.H_{ba}- p_1.N.H_{ba}\right)+\frac{p_1.k_3\,p_2.D.p_2}{(p_1.p_2)^2}p_1.H_{ba}\bigg]\right)\nonumber
 \eeqa
 where we have also used the relation $k_3.H_{ab}F^{ab}=F^{ab}k_3^cH_{abc}=0$. In above equation $\phi_1$ is the polarization of the dilaton which is one. Note that there is no massless open string pole $1/p_1.D.p_1$ which is consistent with field theory since there is no linear dilaton coupling for D$_3$-brane in the Einstein frame. This is unlike the graviton amplitude \reef{ampg} which has such pole and is reproduced in field theory by the pull-back  and by the Taylor expansion of the linear coupling of graviton to the D-brane, \ie $\prt_a X^i h^a{}_i$  and $X^i\prt_ih_a{}^a$ where $h$ is the graviton and $X^i$ is the transverse scalar field on the world volume of D-brane.
 
 Similarly, one can find the dilaton amplitudes corresponding to the components \reef{amp26} and \reef{amp27}. The dilaton amplitude corresponding to \reef{amp26} is
 \beqa
 A_{\phi C^{(2)}F}&\sim&2\phi_1T_3e^{-\phi_0/2}\eps^{abcd}F_{cd}\left(I_{11}\bigg[\frac{p_1.D.p_1}{p_1.p_2}  p_1.V.F^{(3)}_{ba}+\frac{p_1.k_3\,p_1.D.p_1}{(p_1.p_2)^2}p_1.F^{(3)}_{ba}\bigg]\right.\labell{amp28}\\
 &&\left.-I_2\bigg[4p_1.V.F^{(3)}_{ba}+\frac{p_2.D.p_2}{p_1.p_2} \left(2 p_1.F^{(3)}_{ba}- p_1.N.F^{(3)}_{ba}\right)+\frac{p_1.k_3\,p_2.D.p_2}{(p_1.p_2)^2}p_1.F^{(3)}_{ba}\bigg]\right)\nonumber
 \eeqa
 The dilaton amplitude corresponding to \reef{amp27} is
 \beqa
 A_{\phi BFC^{(0)}}&\sim&2\phi_1C_0T_3e^{-\phi_0/2}\eps^{abcd}F_{cd}\left(I_{11}\bigg[\frac{p_1.D.p_1}{p_1.p_2}  p_1.V.H_{ba}+\frac{p_1.k_3\,p_1.D.p_1}{(p_1.p_2)^2}p_1.H_{ba}\bigg]\right.\labell{amp29}\\
 &&\left.-I_2\bigg[4p_1.V.H_{ba}+\frac{p_2.D.p_2}{p_1.p_2} \left(2 p_1.H_{ba}- p_1.N.H_{ba}\right)+\frac{p_1.k_3\,p_2.D.p_2}{(p_1.p_2)^2}p_1.H_{ba}\bigg]\right)\nonumber
 \eeqa
 where the R-R scalar is again constant.
 
 Even though the gravity amplitudes  \reef{amp26}, \reef{amp27} and minus of  \reef{amp25} form an S-dual multiplet, the corresponding dilaton amplitudes  \reef{amp28}, \reef{amp29} and  \reef{ampphi} do not from an S-dual multiplet. This is resulted from the fact that unlike the graviton, the dilaton is not invariant under the S-duality. Therefore, one has to add more amplitudes to find an S-dual multiplet. In the next section we will find this multiplet.
 
 \subsection{Dilaton S-dual multiplet}
 
 To find the S-dual multiplet corresponding to the amplitudes \reef{ampphi}, \reef{amp28} and \reef{amp29},  consider  the variation of the matrix \reef{M} which is given by
\beqa
\delta\cM=\pmatrix{-(e^{-\phi}-C^2e^{\phi})\delta\phi+2C e^{\phi}\delta C& C e^{\phi}\delta\phi+e^{\phi}\delta C\cr 
C e^{\phi}\delta\phi+e^{\phi}\delta C&e^{\phi}\delta \phi}\labell{dM}
\eeqa
This transforms under the $SL(2,R)$ transformation as
\beqa
{\delta\cal M}\rightarrow \Lambda \delta{\cal M}\Lambda ^T
\eeqa
Consider  the case that the variations are the external states, \ie $\delta\phi=\phi_1$ and $\delta C=C_1$, and the dilaton and the axion are the constant background fields $\phi_0$ and $C_0$, respectively. Using this matrix and the doublets \reef{2}, one finds the following S-dual multiplet:
\beqa
(*\cF)^T\delta\cM\cB&=&e^{-\phi_0}\phi_1FB+\phi_1(*F)C^{(2)}+C_0\phi_1(*F)B\nonumber\\
&&+C_1(*F)B-e^{-\phi_0}C_0C_1FB-e^{-\phi_0}C_1FC^{(2)}\labell{dual}
\eeqa
 The amplitudes \reef{ampphi}, \reef{amp28} and \reef{amp29}, correspond to the first, the second and the third  components of the multiplet \reef{dual},  respectively. The amplitude associated with the fourth component is 
\beqa
 A_{C^{(0)}BF}&\sim&2C_1T_3e^{-\phi_0/2}\eps^{abcd}F_{cd}\left(I_{11}\bigg[\frac{p_1.D.p_1}{p_1.p_2}  p_1.V.H_{ba}+\frac{p_1.k_3\,p_1.D.p_1}{(p_1.p_2)^2}p_1.H_{ba}\bigg]\right.\labell{amp30}\\
 &&\left.-I_2\bigg[4p_1.V.H_{ba}+\frac{p_2.D.p_2}{p_1.p_2} \left(2 p_1.H_{ba}- p_1.N.H_{ba}\right)+\frac{p_1.k_3\,p_2.D.p_2}{(p_1.p_2)^2}p_1.H_{ba}\bigg]\right)\nonumber
 \eeqa
 where $C_1$ is the polarization of the R-R scalar which is one.  The amplitude corresponding to the fifth component is 
 \beqa
A_{C^{(0)}BFC^{(0)}}&\sim&-4C_1C_0T_3e^{-3\phi_0/2}F^{ab}\left(I_{11}\bigg[\frac{p_1.D.p_1}{p_1.p_2}  p_1.V.H_{ba}+\frac{p_1.k_3\,p_1.D.p_1}{(p_1.p_2)^2}p_1.H_{ba}\bigg]\right.\labell{amp31}\\
 &&\left.-I_2\bigg[4p_1.V.H_{ba}+\frac{p_2.D.p_2}{p_1.p_2} \left(2 p_1.H_{ba}- p_1.N.H_{ba}\right)+\frac{p_1.k_3\,p_2.D.p_2}{(p_1.p_2)^2}p_1.H_{ba}\bigg]\right)\nonumber
\eeqa
in which one of the R-R scalars is constant. And the amplitude associated with the last component is
\beqa
A_{C^{(0)}C^{(2)}F}&\sim&-4C_1T_3e^{-3\phi_0/2}F^{ab}\left(I_{11}\bigg[\frac{p_1.D.p_1}{p_1.p_2}  p_1.V.F^{(3)}_{ba}+\frac{p_1.k_3\,p_1.D.p_1}{(p_1.p_2)^2}p_1.F^{(3)}_{ba}\bigg]\right.\labell{amp32}\\
 &&\left.-I_2\bigg[4p_1.V.F^{(3)}_{ba}+\frac{p_2.D.p_2}{p_1.p_2} \left(2 p_1.F^{(3)}_{ba}- p_1.N.F^{(3)}_{ba}\right)+\frac{p_1.k_3\,p_2.D.p_2}{(p_1.p_2)^2}p_1.F^{(3)}_{ba}\bigg]\right)\nonumber
\eeqa
The amplitudes \reef{ampphi}, \reef{amp28}, \reef{amp29}, \reef{amp30}, \reef{amp31} and \reef{amp32} form an S-dual multiplet. Since the leading $\alpha'$-order terms of the integrals $I_2$ and $I_{11}$ have the dilaton factor $e^{-\phi_0/2}$, the leading terms of the S-dual multiplet have no extra dilaton factor so they are  invariant under the S-duality. %The extra dilaton factors in all higher order terms may be again dressed with the loops and the nonperturbative effects \cite{Green:1997tv} - \cite{Garousi:2011fc} to become invariant under the full S-duality.
The above results are the S-duality prediction for the S-matrix elements. The explicit calculation of the amplitudes \reef{amp31} and \reef{amp32} needs the correlation function of four spin operators and some world-sheet fermions   \cite{Hartl:2009yf,Hartl:2010ks} which we leave them for future works.  The    amplitude \reef{amp30} is a 5-point function that  needs the correlation function of two spin operators and some fermions.    In the next section we   calculate this amplitude explicitly. 
 
 \section{Testing the multiplets}
 
 The scattering amplitude of one R-R $n$-form, one NS-NS and one gauge boson may be given by the following correlation function:
\beqa
\cA&\sim&<V_{RR}^{(-1/2,-3/2)}(\veps_1^{(n)},p_1)V_{NSNS}^{(0,0)}(\veps_2,p_2)V_{A}^{(0)}(\z_3,k_3)>\labell{amp23}
\eeqa
where the vertex operators are \cite{Billo:1998vr,Garousi:1996ad}
\beqa
V_{RR}^{(-1/2,-3/2)}&\!\!\!\!\!=\!\!\!\!\!&(P_-H_{1(n)}M_p)^{AB}\int d^2z_1:e^{-\phi(z_1)/2}S_A(z_1)e^{ip_1\cdot X}:e^{-3\phi(\bz_1)/2}S_B(\bz_1)e^{ip_1\cdot D\cdot  X}:\nonumber\\
V_{NSNS}^{(0,0)}&\!\!\!\!\!=\!\!\!\!\!&(\veps_2\inn D)_{\mu\nu}\int d^2z_2:(\prt X^{\mu}+ip_2\inn\psi\psi^{\mu})e^{ip_2\cdot X}:(\prt X^{\nu}+ip_2\inn D\inn\psi\psi^{\nu})e^{ip_2\cdot D\cdot X}:\nonumber\\
V_{A}^{(0)}&\!\!\!\!\!=\!\!\!\!\!&(\z_3)_{a}\int dx_3:(\prt X^{a}+2ik_3\inn\psi\psi^{a})e^{2ik_3\cdot X}
\eeqa
where  the indices $A,B,\cdots$ are the Dirac spinor indices and  $P_-=\frac{1}{2}(1-\gamma_{11})$ is the chiral projection operator which makes the calculation of the gamma matrices to be with the full $32\times 32$ Dirac matrices of the ten dimensions. 
 In the R-R vertex operator, $H_{1(n)}$ and $M_p$ are
\beqa
H_{1(n)}&=&\frac{1}{n!}\veps_{1\mu_1\cdots\mu_{n}}\gamma^{\mu_1}\cdots\gamma^{\mu_{n}}\nonumber\\
M_p&=&\frac{\pm 1}{(p+1)!} \eps_{a_0 \cdots a_p} \ga^{a_0} \cdots \ga^{a_p}
\eeqa
where $\eps$ is the volume $(p+1)$-form of the $D_p$-brane and $\veps_1$ is the polarization of the R-R form.

 Using the propagators \reef{wpro}, one can easily calculate the $X$ and $\phi$ correlators in \reef{amp23}. 
To find the correlator of $\psi$, we use the following Wick-like rule for the correlation function involving an arbitrary number of $\psi$'s and two $S$'s \cite{Liu:2001qa,Garousi:2008ge,Garousi:2010bm}:
\beqa
 &&<:S_{A}(z_1):S_{B}(\bz_1):\psi^{\mu_1}
(z_2)\cdots
\psi^{\mu_{n-1}}(z_n):>=\labell{wicklike}\\
&&\frac{1}{2^{n/2}}
\frac{(z_{1\bar{1}})^{n/2-5/4}}
{\sqrt{z_{21}z_{2\bar{1}}}\cdots\sqrt{z_{n1}z_{n\bar{1}}}}\left\{(\gamma^{\mu_{n-1}\cdots\mu_1}
C^{-1})_{AB}+\cP(z_3,z_2)\eta^{\mu_2\mu_1}(\gamma^{\mu_{n-1}\cdots\mu_3}
C^{-1})_{AB}\right.\nonumber\\
&&\left.
+\cP(z_3,z_2)\cP(z_5,z_4)\eta^{\mu_2\mu_1}\eta^{\mu_4\mu_3}
(\gamma^{\mu_{n-1}\cdots\mu_5}
C^{-1})_{AB}+\cdots\pm {\rm perms}\right\}\nonumber\eeqa where  dots mean  sum  over all possible contractions. In above equation, $\gamma^{\mu_{n}...\mu_{1}}$ is the totally antisymmetric combination of the gamma matrices and  $\cP(z_i,z_j)$ is given by the Wick-like contraction
\beqa
\cP(z_i,z_j)\eta^{\mu\nu}&=&\widehat{[\psi^{\mu}(z_i),\psi^{\nu}(z_j)]}=\eta^{\mu\nu}{\frac {z_{i1}z_{j\bar{1}}+z_{j1}z_{i\bar{1}}}{z_{ij}z_{1\bar{1}}}}\labell{ppp}
\eeqa
 Combining the gamma matrices coming from the  correlation \reef{wicklike} with the gamma matrices in the R-R vertex operator, one finds  the amplitude \reef{amp23} has the  following  trace:
 \beqa
T(n,p,m)& =&(H_{1(n)}M_p)^{AB}(\gamma^{\alpha_1\cdots \alpha_m}C^{-1})_{AB}A_{[\alpha_1\cdots \alpha_m]}\labell{relation1}\\
& =&\frac{1}{n!(p+1)!}\veps_{1\nu_1\cdots \nu_{n}}\eps_{a_0\cdots a_p}A_{[\alpha_1\cdots \alpha_m]}\Tr(\gamma^{\nu_1}\cdots \gamma^{\nu_{n}}\gamma^{a_0}\cdots\gamma^{a_p}\gamma^{\alpha_1\cdots \alpha_m})\nonumber
 \eeqa
where $A_{[\alpha_1\cdots \alpha_m]}$ is an antisymmetric combination of the momenta and  the polarizations of the NS-NS field and the gauge boson. 
The trace \reef{relation1} can be evaluated for specific values of $n$ and $p$. Since we are going to test the amplitudes that have been found in the previous sections by S-duality, we consider only the case of $p=3$. 

 \subsection{R-R two-form amplitude}
 
 In this section we calculate the amplitude for the R-R two-form, \ie $n=2$. The trace \reef{relation1} then becomes
 \beqa
 T(2,3,m)
& =&\frac{1}{2!4!}\veps_{1\nu_1\nu_{2}}\eps_{a_0\cdots a_3}A_{[\alpha_1\cdots \alpha_m]}\Tr(\gamma^{\nu_1} \gamma^{\nu_{2}}\gamma^{a_0}\cdots\gamma^{a_3}\gamma^{\alpha_1\cdots \alpha_m})\labell{trace}
 \eeqa
 When  both indices of the R-R potential are along the transverse space,  $m=6$. The trace then becomes
 \beqa
T(2,3,6)&=&-\frac{6!}{2!4!}(\veps_1) _{ij}\epsilon_{a_0a_1a_2a_3}A^{[ija_0a_1a_2a_3]}  
\eeqa  
 There is only one such contribution to the amplitude  \reef{amp23}. So the amplitude  can easily be calculated, \ie
 \beqa
 {\cal A}&\sim&\veps_1^{ij}\eps^{a_0\cdots a_3}(\veps_2.D)_{[ij}(\z_3)_{a_0}(p_2)_{a_1}(p_2.D)_{a_2}(k_3)_{a_3]}J\labell{amp241}
 \eeqa
 where the integral $J$ is 
 \beqa
 J&=&\frac{z_{1\bar{1}}}{z_{12} z_{31} z_{2 \bar{1}} z_{3 \bar{1}} z_{1\bar{2}} z_{\bar{1} \bar{2}}}K
 \eeqa
 Since the momentum and the polarization of the gauge boson are along the world-volume, the transverse index in the antisymmetric combination in \reef{amp241} can not be carried by $\z_3$ and $k_3$. Moreover, the two momenta $p_2$ and $p_2.D$ can not both be contracted with the volume form, so the only possible case is when one indices of the NS-NS polarization is along the brane and the other one is along the transverse space. In this case the result becomes zero when the NS-NS polarization is antisymmetric. For the graviton, it becomes 
 \beqa
 {\cal A}_{C^{(2)}gF}&\sim&T_3\veps_1^{ij}\eps^{a_0\cdots a_3}(\veps_2)_{ia_0}(p_2)_j(p_2)_{a_1}F_{a_2a_3}J\nonumber
 \eeqa
 where we have also normalized the amplitude by the D$_3$-brane tension. Note that the amplitude is consistent with T-duality, so there would be no dilaton factor in the string frame. The above amplitude in the Einstein frame becomes
 \beqa
{\cal A}_{C^{(2)}gF}&\sim&T_3 e^{-\phi_0/2}(*F)^{ab}(p_2)_a(p_2\inn \veps_1\inn\veps_2)_b J\labell{test.1}
 \eeqa
 This amplitude should be compared with the corresponding term in the amplitude \reef{amp26}. We are not going to rewrite the amplitudes in this section in terms of the R-R field strength, so we have to compare the above amplitude with the amplitude predicted by imposing the S-duality on the NS-NS amplitude \reef{amp20}.
 
So we  write the amplitude   \reef{amp20} for the special case that the B-field has polarization only in the transverse space which is
\beqa
A_{gB_{ij}F}&\sim&4T_p p_1.F.\veps _1.\veps _2.p_1 \left(I_1-I_2-I_3\right)\labell{18}
\eeqa
Using the relations \reef{rel.1}, the integrals add up to $I_8$. This integral is
\beqa
I_8&=&\frac{z_{2 \bar{2}} }{z_{12} z_{32} z_{2 \bar{1}} z_{1\bar{2}} z_{3 \bar{2}} z_{\bar{1} \bar{2}}}K
\eeqa
%Since the leading $\alpha'$-order terms in the amplitude \reef{ampg} are one-momentum order, the leading $\alpha'$-order term of the integral $I_8$ is massless pole. 
The S-duality then predicates the following amplitude, in the Einstein frame, for the S-matrix element of one graviton, one R-R two-form and one gauge boson on the world volume of the D$_3$-brane: 
\beqa
A_{gC^{(2)}F}&\sim&4I_8T_3e^{-\phi_0/2} p_1.(*F).\veps _1.\veps _2.p_1 
\eeqa
where $\veps_1$ is the polarization of the graviton and $\veps_2$ is the polarization of the R-R two-form.  The above amplitude  is exactly the amplitude \reef{test.1} in which the labels of the graviton and the R-R two-form are interchanged.

 When one index of the R-R potential is along the world-volume and the other one is along the transverse space,  $m=4$. In this case the trace \reef{trace} becomes
\beqa
T(2,3,4)&=&\veps_1^{a_0}{}_{i}\epsilon_{a_0a_1a_2a_3}A^{[ia_1a_2a_3]} \labell{T23}
\eeqa    
 There are many such terms in the amplitude \reef{amp23}. In fact 
 performing the correlators in the amplitude \reef{amp23}, one finds the following result  in terms of the gauge field strength:
 \beqa
 {\cal A} &\!\!\!\!\sim\!\!\!\!&T_3F_{ ab} \bigg[\left(p_2.D\right)_{\beta _4} \left(p_2\right)_{\beta _3} \left(-4 I_8 (\gamma ^{a \beta _3 \beta _4 \mu _3}C^{-1})_{AB} \left(\veps _2\right)_{ \mu _3}{}^{ b}+J_4 (\gamma ^{a b \beta _3 \beta _4}C^{-1})_{AB} \Tr\left[\veps _2.D\right]\right)\nonumber\\
  &&+(\gamma ^{a b \beta _3 \mu _3}C^{-1})_{AB} \left(\left(p_2.D\right)_{\beta _3} \left(4 J_5 \left(k_3.\veps _2.D\right)_{\mu _3}+2 I_1 \left(p_1.\veps _2.D\right)_{\mu _3}+J_4 \left(p_2.D.\veps _2\right)_{\mu _3}\right.\right. \nonumber\\
  &&\left.\left. -2 I_2 \left(p_1.D.\veps _2.D\right)_{\mu _3}\right)-\left(p_2\right)_{\beta _3}\left(-4 J_5^* \left(k_3.\veps _2\right)_{\mu _3}+2 I_1 \left(p_1.\veps _2\right)_{\mu _3}-2 I_2 \left(p_1.D.\veps _2\right)_{\mu _3}\right.\right. \nonumber\\
  &&\left.\left. +J_4 \left(p_2.D.\veps _2.D\right)_{\mu _3}\right) \right) + p_2.D.p_2 J_4 (\gamma ^{a b \mu _3 \mu _4}C^{-1})_{AB} \left(\veps _2.D\right)_{\mu _3 \mu _4} \nonumber\\
  &&+2  (\gamma ^{a \beta _3 \mu _3 \mu _4}C^{-1})_{AB}  \left(\veps _2.D\right)_{\mu _3 \mu _4} \left(p_2\right)_b(J_0\left(p_2.D\right)_{\beta _3}-J_0^*\left(p_2\right)_{\beta _3} ) 
  %\nonumber\\
 % &&+2  (\gamma ^{a \beta _3 \beta _4\mu _3  }C^{-1})_{AB}  \left(\veps _2.D\right)_{b\mu _3  } \left(p_2\right)_{\beta _3}\left(p_2.D\right)_{\beta _4}(J[7] -J[8]  )  
 \bigg](H_{1(2)}M_3)^{AB}
 \eeqa
 where integrals $J_0,\, J_4,\, J_5$ are given in the appendix. Using the condition that one of the indices of the antisymmetric gamma matrices is transverse, %one can write the above amplitude as
% \beqa
% {\cal A} &\!\!\!\!\sim\!\!\!\!&2T_3F_{ ab} \bigg[\left(p_2\right)_{i} \left(p_2\right)_{c} \left(-4 I_8 (\gamma ^{a di c}C^{-1})_{AB} (\veps _2)_{d}{}^{ b}+J_4 (\gamma ^{a b ic}C^{-1})_{AB} \Tr\left[\veps _2.D\right]\right)\nonumber\\
 % &&+(\gamma ^{a b ic}C^{-1})_{AB} \left(\frac{}{}\left(p_2\right)_{c} \left( \frac{}{}2(J_5-J_5^*) \left(k_3.\veps _2\right)_{i}+2 I_1 \left(p_1.\veps _2\right)_{i}+J_4 \left(p_2.D.\veps _2\right)_{i}\right.\right. \nonumber\\
%  &&\left.\left. -2 I_2 \left(p_1.D.\veps _2\right)_{i}\frac{}{}\right)-\left(p_2\right)_{i}\left( \frac{}{}2(J_5-J_5^*) \left(k_3.\veps _2\right)_{c}+2 I_1 \left(p_1.\veps _2\right)_{c}-2 I_2 \left(p_1.D.\veps _2\right)_{c}\right.\right. \nonumber\\
%  &&\left.\left. +J_4 \left(p_2.D.\veps _2\right)_{c}\frac{}{}\right)+J_4p_2.D.p_2(\veps_2)_{ic}\frac{}{} \right) -4I_8 (\gamma ^{a di c}C^{-1})_{AB} (p_2)^b(p_2)_d(\veps _2)_{ ic} \bigg](H_{1(2)}M_3)^{AB}\nonumber
% \eeqa
% where we have used the relation $J_0-J_0^*=-2I_8$.    
performing the trace over the gamma matrices and using the relations $(*F)_{ab}=\eps_{abcd}F^{cd}/2$ and  $F_{ab}=-\eps_{abcd}(*F)^{cd}/2$, one finds
 \beqa
 {\cal A} &\!\!\!\!\!\sim\!\!\!\!\!&4T_3(\veps_1)_{a_0}{}^i  \bigg[\left(p_2\right)_{i} \left(p_2\right)_{c} \left( I_8 \eps^{a_0a d c} \eps_{abef}(*F)^{ef}  (\veps _2)_{d}{}^{ b}+2J_4 (*F)^{a_0  c} \Tr\left[\veps _2.V\right]\right)+\nonumber\\
  &&\left(\frac{}{}\left(2p_2\right)_{i} \left( I_8\left(k_3.\veps _2\right)_{c}- J \left(p_1.N.\veps _2\right)_{c}    \right)-\left(2p_2\right)_{c}\left(  I_8 \left(k_3.\veps _2\right)_{i}- J \left(p_1.N.\veps _2\right)_{i}  +2J_4(p_2.V\veps_2)_i  \right)\right.\nonumber\\
  &&\left.+2J_4p_2.V.p_2(\veps_2)_{ic}\frac{}{} \right)(*F)^{a_0  c}  +I_8 \eps^{a_0a d c} \eps_{abef}(*F)^{ef} (p_2)^b(p_2)_d(\veps _2)_{ic}  \bigg] \labell{a2}
  \eeqa
where we have also used the relations $I_1-I_2=J_4$ , $I_1+I_2=J$, $J_0-J_0^*=-2I_8$ and $J_5-J_5^*-J_4=-I_8$.  To compare the above amplitude with corresponding amplitude predicted by the S-duality, we have to write both amplitudes either in terms of the matrices $\eta,\, D$ or in terms of the world volume and the transverse matrices $V,\, N$. Since we have specified the indices of the R-R potential in terms of the world volume and transverse indices, we have to write the amplitudes in terms of $V$ and $N$. In above amplitude we have  written the indices in terms of the world volume and the transverse indices.

The amplitude predicted by the S-duality has no term with square of the world volume form, so we have to use  the following identity:
\beqa
\eps^{abcd}\eps^{e}{}_b{}^{fg}=-\left|\matrix{\eta^{ae}& \eta^{af}& \eta^{ag} \cr 
\eta^{ce}& \eta^{cf}& \eta^{cg} \cr
\eta^{de}& \eta^{df}& \eta^{dg} }\right|
 \labell{iden.3} \eeqa 
% \beqa
%\eps_{a}{}^{cde}\eps_{bc}{}^{fg}=-\eta_{ab}(\eta^{df}\eta^{eg}-\eta^{dg}\eta^{ef})+\delta_a^f(\delta_b^d\eta^{eg}-\delta_b^e\eta^{dg})-\delta_a^g(\delta_b^d\eta^{ef}-\delta_b^e\eta^{df})\labell{iden.3}
%\eeqa
to write the first   and the last terms in \reef{a2} as
\beqa
\eps^{a_0a d c} \eps_{abef}(*F)^{ef}  (\veps _2)_{d}{}^{ b}&=&2(\veps_2.*F)^{a_0c}-2(\veps_2.*F)^{ca_0}-2(*F)^{a_0c}\Tr[\veps_2.V]\nonumber\\
\eps^{a_0a d c} \eps_{abef}(*F)^{ef} (p_2)^b(p_2)_d&=&2p_2^{a_0}(p_2.*F)^c-2p_2^c(p_2.*F)^{a_0}-2p_2.V.p_2(*F)^{a_0c}
\eeqa

 Now one can compare the amplitude \reef{a2} with the amplitude \reef{amp26} predicted by the S-duality. The amplitude \reef{amp26} in terms of the RR potential is the same as \reef{amp20} in which the gauge field strength $F$ is replaced by $(*F)$, the $\veps_2$ is the polarization of the RR potential and $\veps_1$ is the polarization of the graviton. For the case that the RR potential has one transverse and one tangent indices, the amplitude in terms of the matrices $V,\, N$ becomes
 \beqa
 {\cal A} &\sim&T_3\bigg[4 p_ 1. *F.\veps _ 1. N.\veps _ 2. V.p_ 1J + 
 4 p_ 1. *F.\veps _ 1. V.\veps _ 2. N.p_ 
  1J \nonumber\\&&+ 
 4 p_ 1. *F.\veps _ 2. N.\veps _ 1. N.p_ 
  2I_8 + 
 4 p_ 1. *F.\veps _ 2. N.\veps _ 1. V.p_ 
  2J\nonumber\\&&- 
 4 p_ 1. N.\veps _ 2. *F.\veps _ 1. V.p_ 
  2J - 
 4 p_ 1. N.\veps _ 2. *F.\veps _ 1. N.p_ 
  2I_8 \nonumber\\&&+ 
 8 p_ 1. *F.\veps _ 2. N.\veps _ 1. V.p_ 
  1(I_7+I_9) \nonumber\\&&- 
 4\left (I_ 7 + I_ 9 \right)\left (p_ 1. *F.\veps _ 2. N.p_ 
     1\Tr\left[\veps _ 1.V \right] + 
    p_ 1.V.p_ 1\Tr\left[*F.\veps _ 1. N.\veps _ 2 \right] \right)\bigg]\labell{a3}
 \eeqa
where we have also used the relations $4 I_ 2 + 2 I_ 3 + I_ 4 - I_ {10}=2J$, $2 I_ 2 + I_ 3 - I_ 5 + 2 I_ {11}=2(I_7+I_9)$, $I_{10}-I_4=-2I_8$ and $I_ 1 + I_ 2=J$. Using the explicit form of the integrals, one also finds 
\beqa
I_8(p_1,p_2,k_3)-J_4(p_1,p_2,k_3)=I_7(p_2,p_1,k_3)+I_9(p_2,p_1,k_3)&\!\!;\!\!&J(p_1,p_2,k_3)=I_8(p_2,p_1,k_3)\nonumber
\eeqa
Using the above relations and using   the conservation of momentum to rewrite $k_3.\veps_2$ in \reef{a2} as $-p_2.V.\veps_2-p_1.V.\veps_2 $,
%\beqa
%k_3.\veps_2=-p_2.V.\veps_2-p_1.V.\veps_2 
%\eeqa
 one finds  the amplitude \reef{a2} is exactly the amplitude  \reef{a3} in which the labels of the graviton and the R-R two-form are interchanged.
%\beqa
%4I_1(p_1,p_2,p_3)-2J_4(p_1,p_2,p_3)=I_4(p_2,p_1,p_3)-I_{10}(p_2,p_1,p_3)
%\eeqa

 When  both indicies of the R-R polarization are along the world-volume,  the value of $m$ in the trace \reef{trace} can be $m=2$ and $m=4$. However, the trace for the $m=4$ case has the following structure:
\beqa
T(2,3,4)&=&4\veps_1^{a_0}{}_{b}\epsilon_{a_0a_1a_2a_3}A^{[ba_1a_2a_3]} \labell{T234}
\eeqa
which can be shown it is zero by writing the RR polarization as $(\veps_1)_{a_0b}=-(*\veps_1)_{\alpha\beta}\epsilon_{a_0b}{}^{\alpha\beta}/2$ and using the identity \reef{iden.3}.
  So the trace is nonzero only for $m=2$ which is
 \beqa
T(2,3,2)&=&\frac{1}{2!}\veps_1^{a_0a_1} \epsilon_{a_0a_1a_2a_3}A^{[ a_2a_3]} \labell{T2}\\
&=&-(*\veps_1)_{a_2a_3}A^{[ a_2a_3]}\nonumber
\eeqa
In this case also there are many such terms in the amplitude \reef{amp23}. Performing the correlators in \reef{amp23}, one finds the following result in terms of the gauge field strength:
\beqa
{\cal A}&\!\!\!\!\!\sim\!\!\!\!\!& -8 p_2.F.\veps _2.*\veps _1.p_2 J_6-4 p_1.\veps _2.F.*\veps _1.p_2 J_{11}-4 p_1.D.\veps _2.F.*\veps _1.p_2 J_{13}\nonumber\\
&&-8 k_3.\veps _2.F.*\veps _1.p_2 J_{15}-4 p_2.F.*\veps _1.\veps _2.p_1 J_{16}-4 p_2.F.*\veps _1.\veps _2.D.p_1 J_{17}\nonumber\\
&&+4 k_3.\veps _2.p_1 J_7 \Tr\left[F.*\veps _1\right]-4 k_3.\veps _2.D.p_1 J_8 \Tr\left[F.*\veps _1\right]+p_1.\veps _2.p_1 J_{10} \Tr\left[F.*\veps _1\right]\nonumber\\
&&-p_1.D.\veps _2.D.p_1 J_{12} \Tr\left[F.*\veps _1\right]-4 k_3.\veps _2.k_3 J_{15} \Tr\left[F.*\veps _1\right]-2 p_1.\veps _2.D.p_1 J_{15} \Tr\left[F.*\veps _1\right]\nonumber\\
&&-4 p_2.F.*\veps _1.p_2 J_9 \Tr\left[\veps _2.D\right]+p_2.D.p_2 J_{19} \Tr\left[F.*\veps _1\right] \Tr\left[\veps _2.D\right]+8 p_2.F.*\veps _1.\veps _2.k_3 J_{18}\nonumber\\
&&+4 J_{20} \Tr\left[F.*\veps _1\right] \Tr\left[\veps _2.D\right]-4 p_2.D.p_2 J_9 \Tr\left[F.*\veps _1.\veps _2\right]+8 p_2.F.\veps _2.*\veps _1.p_2 J_{14}\labell{a4}
\eeqa
where $J_6,\cdots, J_{20}$ are some integrals that appear in the Appendix. While it was straightforward to write all above terms in terms of gauge field strength, the first term is in fact in the form of 
\beqa
-\frac{1}{2}  I \z .\eps _2.*\veps _1.p_2+ J_6 p_1.\z  k_3.\veps _2.*\veps _1.p_2
\eeqa
where $\z$ is the polarization of the gauge boson and $I$ is an integral which appears in the Appendix. The gauge symmetry Ward identity dictates that the integrals $I$ and $J_6$ must satisfy the relation $I=2p_1.k_3 J_6$ which may be verified using   by part integration. Using this relation, the above terms can be written as the first term in \reef{a4}. As a check of our calculation, we have imposed the Ward identity corresponding to the graviton. We have found the  relations $J_{16}=2 J_9-J_{11}$, $J_{17}=2 J_9-J_{13}$ and $J_{18}=-2 J_9+J_{15}$ 
%between the integrals:
%\beqa
% J_{16}=2 J_9-J_{11},\,\,\,J_{17}=2 J_9-J_{13},\,\,\,J_{18}=-2 J_9+J_{15}
%\eeqa
which can easily be checked with the explicit form of the integrals. This Ward identity gives also some other relations in which the Mandelstam variables appear, \eg
\beqa
J_{20}=\frac{1}{4} \left(-p_1.p_2 J_7-p_1.D.p_2 J_7-p_2.D.p_2 J_7+p_1.p_2 J_{10}-p_1.D.p_2 J_{15}-p_2.D.p_2 J_{19}\right)
\eeqa
Using this relation, one finds that the amplitude \reef{a4} has no tachyon pole as expected. Note that  the explicit form of integrals $J_{19},\, J_{20}$ show that they have tachyon poles, however, the combination $4J_{20}+p_2.D.p_2 J_{19}$ which appears in the amplitude \reef{a4}, can be written in terms of tachyon-free integrals $J_{7},\, J_{10}, \, J_{15}$ via the above relation. Similarly, the integrals  $J_6$ and $J_{14}$ have tachyon pole,  however, the combination $J_6-J_{14}$ that appears in the amplitude \reef{a4} has no tachyon pole. There is in fact the relation
$J_6-J_{14}=J_{18}$ where the integral $J_{18}$ has no tachyon pole.

Now to compare the amplitude \reef{a4} with the corresponding terms in the amplitude \reef{amp26}, we have to write the amplitude \reef{a4} in terms of $*F$ and $\veps_1$. To this end, we use the identity \reef{iden.3} to write the terms in \reef{a4} that have structure $A.F.*\veps_1.B$ as
\beqa
A.F.*\veps_1.B&=&A.\veps _1.*F.B-\frac{1}{2} A.B \,\Tr\left[*F.\veps _1\right]
\eeqa
where $A,B$ are two arbitrary vectors. The above identity can be used to rewrite all terms in \reef{a4} in terms of $*F$ and $\veps_1$. However, the polarization of graviton $\veps_2$ in the first  and in the last terms appear  between $F$ and $*\veps_1$. To rewrite them in terms of $*F$ and $\veps_1$ we have to use  the following identity:
\beqa
\eps^{abcd}\eps^{efgh}=-\left|\matrix{\eta^{ae}& \eta^{af}& \eta^{ag}&\eta^{ah}\cr 
\eta^{be}& \eta^{bf}& \eta^{bg}&\eta^{bh}\cr
\eta^{ce}& \eta^{cf}& \eta^{cg}&\eta^{ch}\cr
\eta^{de}& \eta^{df}& \eta^{dg}&\eta^{dh}}\right|
 \labell{zero2n} 
 \eeqa 
 to find the relation
 \beqa
 p_2.F.\veps _2.*\veps _1.p_2&=&p_2.*F.\veps _1.\veps _2.V.p_2+p_2.*F.\veps _2.\veps _1.p_2+p_2.V.\veps _2.*F.\veps _1.p_2\nonumber\\
 &&-\frac{1}{2} p_2.V.\veps _2.V.p_2 \Tr\left[*F.\veps _1\right]-\frac{1}{2} p_2.*F.\veps _1.p_2 \Tr\left[\veps _2.D\right]\nonumber\\
 &&+\frac{1}{4} p_2.V.p_2 \Tr\left[*F.\veps _1\right] \Tr\left[\veps _2.D\right]-p_2.V.p_2 \Tr\left[*F.\veps _2.\veps _1\right]
 \eeqa
Note that the R-R polarization $\veps_1$ has only world volume indices.  Using the above relations,  the amplitude \reef{a4} in terms of $*F$ and $\veps_1$, and in terms of the matrices $V,\, N$ becomes
 \beqa
{\cal A}&\!\!\!\!\!\sim\!\!\!\!\!& 4 k_3.\veps _1.*F.\veps _2.N.p_1 \left(J_{11}-J_{13}\right)- p_1.N.\veps _2.N.p_1 (-J_{10} +J_{12}-2J_{15})\Tr\left[*F.\veps _1\right]\nonumber\\
&&-4 p_2.V.\veps _2.\veps _1.*F.p_2 \left(2 J_6+J_{11}+J_{13}-2 J_{14}\right)-8 p_2.*F.\veps _2.\veps _1.k_3 \left(-J_6+J_{14}\right)\nonumber\\
&&-4 k_3.\veps _1.*F.\veps _2.V.p_2 \left(-2 J_6-4 J_9+J_{11}+J_{13}+2 J_{14}\right)+4 p_1.N.\veps _2.\veps _1.*F.p_2 \left(J_{11}-J_{13}\right)\nonumber\\
&&-4 k_3.\veps _1.*F.\veps _2.k_3 \left(J_{11}+J_{13}-2 J_{15}\right)-4 k_3.\veps _2.\veps _1.*F.p_2 \left(J_{11}+J_{13}-2 J_{15}\right)\nonumber\\
&& -2 p_1.N.\veps _2.V.p_2 \left(J_{10}+J_{12}\right) \Tr\left[*F.\veps _1\right]-2k_3.\veps _2.N.p_1 \left(-2 \left(J_7+J_8\right)+J_{10}+ J_{12}\right) \Tr\left[*F.\veps _1\right]\nonumber\\
&&-p_2.V.\veps _2.V.p_2 \left(-4 J_6-8 J_9-J_{10}+J_{12}+4 J_{14}+2 J_{15}\right) \Tr\left[*F.\veps _1\right]\nonumber\\
&&-k_3.\veps _2.k_3 \left(4 J_7-4 J_8-J_{10}+J_{12}+6 J_{15}\right) \Tr\left[*F.\veps _1\right]\nonumber\\
&&-2 k_3.\veps _2.V.p_2 \left(-J_{10}+J_{12}+2 \left(J_7-J_8+J_{15}\right)\right) \Tr\left[*F.\veps _1\right]\nonumber\\
&&+8 p_2.*F.\veps _1.k_3 \left(-J_6-J_9+J_{14}\right) \Tr\left[\veps _2.V\right]\nonumber\\
&&+2[2 p_2.k_3 \left(J_7+J_{15}\right)+p_1.p_2 \left(J_{10}+J_{15}\right)+2p_2.V.p_2 \left(-J_6-2 J_9+J_{14}+J_{15}\right)]\nonumber\\
&&\times \Tr\left[*F.\veps _1\right] \Tr\left[\veps _2.V\right] -8 p_2.V.p_2 \left(-J_6-J_9+J_{14}\right) \Tr\left[*F.\veps _2.\veps _1\right]\labell{a5}
 \eeqa
 where we have also used the on-shell relation $\veps_1.V.p_1=0$ because we assumed the R-R potential has only world volume indices. 
 
 We can now compare the above amplitude  with the amplitude \reef{amp26}. This amplitude  in terms of the RR potential is the same as \reef{amp20} in which the gauge field strength $F$ is replaced by $(*F)$, the $\veps_2$ is the polarization of the RR potential and $\veps_1$ is the polarization of the graviton. For the case that the RR potential has only world volume  indices, the amplitude  becomes
\beqa
{\cal A}&\!\!\!\!\!\sim\!\!\!\!\!& -4 p_1.*F.\veps _1.\veps _2.k_3 \left(I_1-I_2+I_3\right) +2 p_1.*F.\veps _2.\veps _1.N.p_2 \left(I_4+I_{10}\right) \nonumber\\
&&+2 k_3.\veps _2.*F.\veps _1.k_3 \left(4 I_2-2 I_3-I_4-I_{10}\right)+2 p_1.*F.\veps _2.\veps _1.k_3 \left(4 I_2-2 I_3-I_4-I_{10}\right)\nonumber\\
&&+2 p_1.V.\veps _1.*F.\veps _2.k_3 \left(4 I_2-2 I_3-I_4-I_{10}\right)  +2 k_3.\veps _2.*F.\veps _1.N.p_2 \left(I_4+I_{10}\right)\nonumber\\
&& -2 p_1.*F.\veps _2.\veps _1.V.p_1 \left(I_4+2 I_5+I_{10}+4 I_{11}\right)+2 p_1.V.\veps _1.V.p_1 I_8 \Tr\left[*F.\veps _2\right]\nonumber\\
&&+2 p_2.N.\veps _1.N.p_2 I_8 \Tr\left[*F.\veps _2\right]-4 k_3.\veps _1.N.p_2 \left(I_8+I_{10}\right) \Tr\left[*F.\veps _2\right]\nonumber\\
&&+2 k_3.\veps _1.k_3 \left(-4 I_2+I_8+2 I_{10}\right) \Tr\left[*F.\veps _2\right]-4 p_1.V.\veps _1.N.p_2 \left(I_8-2 I_{11}\right) \Tr\left[*F.\veps _2\right]\nonumber\\
&&+4 p_1.V.\veps _1.k_3 \left(-2 I_6+I_8+I_{10}-2 I_{11}\right) \Tr\left[*F.\veps _2\right]+4 p_1.*F.\veps _2.k_3 \left(I_7-I_9\right) \Tr\left[\veps _1.V\right]\nonumber\\
&&+4 \left(p_1.k_3 I_2-\left(p_1.k_3+p_1.p_2+2p_1.V.p_1\right) I_{11}\right) \Tr\left[*F.\veps _2\right] \Tr\left[\veps _1.V\right] \nonumber\\
&& +4 p_1.V.p_1 \left(-I_7+I_9\right) \Tr\left[*F.\veps _1.\veps _2\right]\labell{a6}
 \eeqa  
 In order to have equality between the above amplitude and \reef{a5}, there should be the following relation between the integrals in \reef{a5}:
 \beqa
 -J_6-2J_9+J_{14}+J_{15}&=&0,\,\, -4 J_6-8J_9-J_{10}+J_{12}+4J_{14}+6J_{15}=0\nonumber\\
 -J_{10}+J_{12}+2J_{15}&=&0
 \eeqa
 which are true using the explicit form of the integrals. Using the integrals in the Appendix, one finds the following relations between the integrals in amplitude \reef{a4} and the integrals in \reef{a6}:
 \beqa
 (J_{11}-J_{13})(p_1,p_2,k_3)&=&(I_4+I_{10})(p_2,p_1,k_3)\nonumber\\
 (2J_6+J_{11}+J_{13}-2J_{14})(p_1,p_2,k_3)&=&(I_4+2I_5+I_{10}+4I_{11})(p_2,p_1,k_3)\nonumber\\
 (-J_6+J_{14})(p_1,p_2,k_3)&=&(I_1-I_2+I_3)(p_2,p_1,k_3)\nonumber\\
 (2J_6+4J_9-J_{11}-J_{13}-2J_{14})(p_1,p_2,k_3)&=&(4I_2-2I_3-I_4-I_{10})(p_2,p_1,k_3)\nonumber\\
 (J_{11}+J_{13}-2J_{15})(p_1,p_2,k_3)&=&(-4I_2+2I_3+I_4+I_{10})(p_2,p_1,k_3)\nonumber\\
 (J_{10}+J_{12})(p_1,p_2,k_3)&=&4(I_8-2I_{11})(p_2,p_1,k_3)\nonumber\\
 (-2J_7-2J_8+J_{10}+J_{12})(p_1,p_2,k_3)&=&4(I_8+I_{10})(p_2,p_1,k_3)\nonumber\\
 (4J_{7}-4J_8-J_{10}+I_{12}+6J_{15})(p_1,p_2,k_3)&=&4(4I_2-I_8-2I_{10})(p_2,p_1,k_3)\nonumber\\
 (-J_{10}+J_{12}+2J_7-2J_8+2J_{15})(p_1,p_2,k_3)&=&4(2I_6-I_8-I_{10}+2I_{11})(p_2,p_1,k_3)\nonumber\\
 (J_6+J_9-J_{14})(p_1,p_2,k_3)&=&(-I_7+I_9)(p_2,p_1,k_3)
 \eeqa
 One then finds exact agreement between all terms in the two amplitudes except the term which has structure $ \Tr\left[*F.\veps _2\right] \Tr\left[\veps _1.V\right] $. The equality of this term gives the following relation between the integrals:
 \beqa
 &&[2 p_2.k_3 \left(J_7+J_{15}\right)+p_1.p_2 \left(J_{10}+J_{15}\right) ](p_1,p_2,p_3)\nonumber\\
 &&=4 [p_1.k_3 I_2-\left(p_1.k_3+p_1.p_2+2p_1.V.p_1\right) I_{11}] (p_2,p_1,p_3)
 \eeqa
 Using the relations $J_7+J_{15}-2I_{11}+2I_2=2(2I_2+I_3)$ and $J_{10}+J_{15}=4I_1$, one can write the above relation as
 \beqa
 p_2.k_3(2I_2+I_3)+p_1.p_2(I_1+I_2)+p_2.D.p_2I_2=0
 \eeqa
 which is the sum of the two relations in \reef{rel.2} in which the labels  1 and 2 are interchanged. Note that $I_1+I_{2}=J$, $I_3+I_2-I_1=-I_8$ and $(2I_2+I_3)(p_1,p_2,k_3)=-(2I_2+I_3)(p_2,p_1,k_3)$. Therefore, we have found precise agreement between the amplitude \reef{amp26} which is predicted by the S-duality and the explicit calculations.

 \subsection{R-R scalar amplitude}

 In this section we are going to compare the amplitude \reef{amp30} with explicit calculation, so we calculate the amplitude \reef{amp23} for the R-R scalar, \ie $n=0$. The trace \reef{relation1} for the D$_3$-brane then becomes
 \beqa
 T(0,3,m)
& =&\frac{1}{4!}C_1\eps_{a_0\cdots a_3}A_{[\alpha_1\cdots \alpha_m]}\Tr(\gamma^{a_0}\cdots\gamma^{a_3}\gamma^{\alpha_1\cdots \alpha_m})\nonumber
 \eeqa
which is nonzero only for $m=4$. The amplitude for B-field  becomes
\beqa
{\cal A}&\sim&-\frac{1}{4}T_3C_1F_{ab}\bigg[-2 I_8 \epsilon ^{a c d e} \left(p_2\right)^b \left(p_2\right)_e \left(\veps _2\right)_{cd}\labell{amp24}\\
&&+\epsilon ^{a b c d} \left( [J_3 \left(p_1.\veps _2\right)_d + 2 J_2 \left(k_3.\veps _2\right)_d+J_1 \left(p_1.D.\veps _2\right)_d] \left(p_2\right)_c+\frac{J_4}{2} p_2.D.p_2 \left(\veps _2\right)_{cd}\right)\bigg]\nonumber
\eeqa 
 where integral 
  $I_8$ is one of the integrals that appear in \reef{amp20}, and the integrals $J_1,\cdots, J_4$ appear in the Appendix.  
  
  One can check that  integrals in \reef{amp24} satisfy the following equations:
 \beqa
 J_4-J_3=-2 I_1,\,\,J_3-J_1/3=8I_1/3,\,\,I_8+J_2-2 J_3=-4I_1\labell{rel.3}
 \eeqa
where $I_1$ is the integral that appears in \reef{apen}. Moreover, the amplitude  \reef{amp24} satisfies the Ward identity associated with the B-field if the integrals satisfy the following equation as well:
\beqa
(J_2+J_3-2 J_4) p_1.p_2+(J_2-J_4) p_1.D.p_1+(J_1+J_2-2 J_4) p_1.D.p_2=0
\eeqa
Using the above four equations, one can find $J_1,\cdots, J_4$ in terms of $I_1$ and $I_8$. Then witting $I_1,\,I_8$ in terms of $I_2$ and $I_{11}$ using the relations \reef{rel.1} and \reef{rel.2}, one can write the amplitude \reef{amp24} in terms of $I_2$ and $I_{11}$. The result in the Einstein frame is
\beqa
{\cal A}_{C^{(0)}BF} &\sim&-T_3C_1F_{ab}e^{-\phi_0/2}\bigg[Q_1 \epsilon ^{a c d e} \left(p_2\right)^b \left(p_2\right)_e \left(\veps _2\right)_{cd}\labell{amp34}\\
&&+\epsilon ^{a b c d} \left(\frac{}{} [Q_2 \left(p_1.\veps _2\right)_d + Q_3 \left(k_3.\veps _2\right)_d+Q_4 \left(p_2.D.\veps _2\right)_d] \left(p_2\right)_c+Q_5 p_2.D.p_2 \left(\veps _2\right)_{cd}\right)\bigg]\nonumber
\eeqa
where the integrals $Q_1,\cdots, Q_5$ are
\beqa
Q_1&=&\frac{I_{11} \left(-2 p_1.k_3+2 p_1.p_2\right)p_1.D.p_1+I_2 \left(2 p_1.k_3\right)p_2.D.p_2 }{4(p_1.p_2)^2}\nonumber\\
Q_2&=&-\frac{I_2\,p_2.D.p_2}{2p_1.p_2}+\frac{ I_{11}\left(2 p_1.k_3\right)p_1.D.p_1-I_2 \left(2 p_1.k_3\right)p_2.D.p_2}{4(p_1.p_2)^2}\nonumber\\
Q_3&=&-\frac{I_{2}\,p_2.D.p_2}{2p_1.p_2}+\frac{I_{11}\,p_1.D.p_1}{2p_1.p_2}\nonumber\\
Q_4&=&I_2+\frac{I_{2}\,p_2.D.p_2}{4p_1.p_2}-\frac{I_{11} \left(2 p_1.k_3\right)p_1.D.p_1 -I_{2} \left(2 p_1.k_3\right)p_2.D.p_2}{8(p_1.p_2)^2}\nonumber\\
Q_5&=&-\frac{I_2}{4}-\frac{I_2\,p_2.D.p_2}{8p_1.p_2}+\frac{I_{11} \left(2 p_1.k_3\right)p_1.D.p_1-I_2 \left(2 p_1.k_3\right)p_2.D.p_2 }{16(p_1.p_2)^2} 
\eeqa
The scattering amplitude of one R-R $n$-form, one B-field and one gauge boson on the world-volume of D$_p$-brane has been also calculated in \cite{Becker:2011ar} (see eq.(3.20) in \cite{Becker:2011ar}). The integrals in that amplitude have been calculated up to $(\alpha')^2$ order. Using the expansion \reef{expand}, one finds agreement between \reef{amp34} and  the amplitude  in  \cite{Becker:2011ar} for $p=3$ and $n=0$. 

Now using the  identity \reef{iden.3},
%\beqa
%\eps_{a}{}^{cde}\eps_{bc}{}^{fg}=-\eta_{ab}(\eta^{df}\eta^{eg}-\eta^{dg}\eta^{ef})+\delta_a^f(\delta_b^d\eta^{eg}-\delta_b^e\eta^{dg})-\delta_a^g(\delta_b^d\eta^{ef}-\delta_b^e\eta^{df})\labell{iden.3}
%\eeqa
%and the relations $(*F)_{ab}=\eps_{abcd}F^{cd}/2$ and  $F_{ab}=-\eps_{abcd}(*F)^{cd}/2$, 
one can rewrite the amplitude \reef{amp34} in terms of $*F$ which can then be compared with the amplitude \reef{amp30}. The difference between the the  amplitude \reef{amp34} and the amplitude \reef{amp30}  becomes
\beqa
\frac{C_1e^{-\phi_0/2} \left(4 I_2 p_1.k_3+2 I_2 p_1.p_2+(I_2-I_{11}) p_1.D.p_1\right)}{p_1.p_2}\bigg[2 p_1.(*F).\epsilon _2.k_3-p_1.k_3\Tr\left[*F.\epsilon _2\right]\bigg]\nonumber
\eeqa
The expression inside the bracket is $(*F)^{ab}k_3^cH_{abc} $. Writing $H_{abc}=-\eps_{abcd}(*H)^d$, one finds
\beqa
(*F)^{ab}k_3^cH_{abc}&=&2F_{cd}k_3^c(*H)^d=0\labell{iden.2}
\eeqa
where we have used  the on-shell condition $k_3.F_d=0$. This ends our illustration of the precise consistency between the amplitudes   \reef{amp30} found from imposing the S-dual Ward identity, and the explicit calculations.

\section{Low energy couplings}

In this section we find the low energy couplings corresponding to the dilaton S-dual multiplet. Replacing the expansion \reef{expand} in \reef{ampphi}, one finds the following expansion:
\beqa
A_{\phi BF}&\!\!\!\!\!=\!\!\!\!\!&\phi_1T_3e^{-3\phi_0/2}F^{ab}\bigg[-16\frac{ e^{\phi_0/2}p_1.V.H_{ba}}{p_2.D.p_2}\left(1+\frac{2\pi^2}{3}(p_2.k_3)^2e^{-\phi_0}\right)-\frac{4e^{\phi_0/2}p_1.H_{ba}}{p_1.p_2}\nonumber\\
&&-\frac{4}{3} \pi ^2 e^{-\phi_0}[\left(2 p_1.N.p_2-3p_2.k_3- p_2.V.p_2\right) p_2.V.H_{ba}-p_2.V.p_2\,p_1.N.H_{ba}] +\cdots\bigg]\labell{pole}
\eeqa
where dots represent the higher order terms. There is no $\alpha'^2$-order term in the massless closed string pole. This is consistent with the fact that there is no $\alpha'^2$ correction to the supergravity. The $(\alpha')^0$-order massless poles should be reproduce by the supergravity couplings in the bulk and  the DBI couplings on the brane.
 Using the following standard coupling in the type IIB supergravity in the Einstein frame \cite{Becker:2007zj}:
\beqa
\int d^{10}x \sqrt{-g}{\cal H}_{\mu\nu\rho}^T{\cal M}{\cal H}^{\mu\nu\rho}\labell{FCH}
\eeqa
where ${\cal H}=d{\cal B}$,
and the standard brane coupling $T_3B^{ab}F_{ab}e^{-\phi_0}$ resulting from expansion of the DBI action
\beqa
S_{DBI}&=&-T_{3}\int d^4x\sqrt{-\det(g_{ab}+e^{-\phi/2}(B_{ab}+2\pi\alpha' F_{ab}))}
\eeqa
 one finds the massless closed string pole in \reef{pole}. 
 The leading term of the massless open string pole in \reef{pole} is reproduced by the  contact term $T_3\phi B^{ab}F_{ab}e^{-\phi_0}$ and by the massless pole  resulting from the brane couplings $T_3\phi F^{ab}F_{ab}e^{-\phi_0}$ and $T_3B^{ab}F_{ab}e^{-\phi_0}$. 
 
 The $\alpha'^2$-order term of the massless open string pole must be reproduced by the higher derivative of the above  couplings. While the  coupling $T_3B^{ab}F_{ab}e^{-\phi_0}$ has no higher derivative correction, the coupling  $T_3\phi F^{ab}F_{ab}e^{-\phi_0}$ has higher derivative correction which is given by the $\alpha'$ expansion of the scattering amplitude of one dilaton and two gauge fields \cite{Hashimoto:1996kf,Garousi:1998fg}. In the Einstein frame, it is
\beqa
A_{\phi FF}&\sim& T_3e^{-\phi_0}\phi_1F_{ab}F^{ab}\frac{\Gamma(1-2te^{-\phi_0/2})}{[\Gamma(1-te^{-\phi_0/2})]^2}\nonumber\\
&\sim&T_3e^{-\phi_0}\phi_1F_{ab}F^{ab}\left(1+\frac{\pi^2}{6}t^2e^{-\phi_0}+2t^3\z(3)e^{-3\phi_0/2}+\cdots\right)\labell{vert}
\eeqa
where the Mandelstam variable is $t=-2k_2.k_3$. Since the structure of the $(\alpha')^0$-order  and the $(\alpha')^2$-order  terms of the massless open string pole in \reef{pole} are the same, one concludes that the higher derivative of the contact term $T_3\phi B^{ab}F_{ab}e^{-\phi_0}$ must be the same as the higher derivative of the coupling $T_3\phi F^{ab}F_{ab}e^{-\phi_0}$ which can be found from \reef{vert} to be $2\pi^2T_3\phi \prt^c\prt^d F^{ab}\prt_c\prt_dF_{ab}e^{-2\phi_0}/3$. Hence the higher derivative of $T_3\phi B^{ab}F_{ab}e^{-\phi_0}$ must be   $2\pi^2T_3\phi \prt^c\prt^dB^{ab}\prt_c\prt_dF_{ab}e^{-2\phi_0}/3$. These higher derivative terms then reproduce the $\alpha'^2$ terms in the massless pole in \reef{pole}.  The other terms in \reef{pole} are only contact terms. 

Therefore, the contact terms of one dilaton, one B-field and one gauge boson  in the momentum space and in the Einstein frame are
\beqa
L&=&2\pi T_3e^{-\phi_0} B_{ab}F^{ab}\phi+\frac{4\pi^3}{3}T_3e^{-2\phi_0} \bigg[(p_2.k_3)^2B_{ab}F^{ab}\phi\labell{contact}\\
&&+\alpha\,F^{ab}\phi\left(\frac{}{}\left(2 p_1.N.p_2-3p_2.k_3- p_2.V.p_2\right) p_2.V.H_{ba}-p_2.V.p_2\,p_1.N.H_{ba}\right)\bigg]+\cdots\nonumber
\eeqa
where $L$ is the D$_3$-brane action in the momentum space. There is  a delta function imposing the conservation of momentum along the brane and the mesure $\int dp_i$ for each field which we have dropped. The dots refer to the other couplings and to the higher momentum of the above couplings. In above action, $p_1$ is the momentum of dilaton, $p_2$ is the momentum of B-field and $k_3$ is the momentum of the gauge field. The first term is coming from the expansion of DBI action. The constant factor $\alpha$ has been added because we have not fix the overall numeric of the string theory amplitudes.  It is important to note that the presence of the second term in \reef{contact} depends on the presence of similar term at the leading order.

Using the same steps for all other components \reef{amp28},\reef{amp29}, \reef{amp30}, \reef{amp31} and \reef{amp32}, one finds the terms corresponding to the second line of \reef{contact} for all these components. However, the terms corresponding to the second term in the first line of \reef{contact} exist only for those components which have couplings at $(\alpha')^0$-order as well. The D-brane action at order $(\alpha')^0$ has only two components of the multiplet \reef{dual}, \ie the first term which is coming from the DBI action and the fourth term which is coming from  the Chern-Simons part of the  D$_3$-brane action. Therefore, the contact terms corresponding to the dilaton S-dual multiplet is given by the following action in the momentum space:
\beqa
{\cal L}&\!\!\!\!\!=\!\!\!\!\!&2\pi T_3B_{ab}[e^{-\phi_0} F^{ab}\phi+(*F)^{ab}C]+\frac{4\pi^3}{3}T_3e^{-\phi_0} \bigg[(p_2.k_3)^2B_{ab}[e^{-\phi_0} F^{ab}\phi+(*F)^{ab}C]\labell{final}\\
&&+\alpha\, (*\cF^T)^{ab}\delta\cM\left(\frac{}{}\left(2 p_1.N.p_2-3p_2.k_3- p_2.V.p_2\right) p_2.V.{\cal H}_{ba}-p_2.V.p_2\,p_1.N.{\cal H}_{ba}\right)\bigg]+\cdots\nonumber
\eeqa 
Each term in the second line represent six different terms according to the expansion \reef{dual}. Unlike the terms in the first line,  the terms in the second line are  invariant under the S-duality, apart from the overall dilaton factor. This factor may be extended to the regularized non-holomorphi Eisenstein series $E_1(\phi_0,C_0)$ after including the loops and the nonperturbative effects \cite{Bachas:1999um,Basu:2008gt,Garousi:2011fc}. To make the couplings to be invariant under the gauge transformation $\delta B=d\Lambda,\, \delta A=-\Lambda/4\pi$, one has to replace $4\pi F$ and $B$ with $B+4\pi F$.

Finally, let us compare the couplings \reef{contact} with the couplings that have been found in \cite{Garousi:2011ut,Becker:2011ar} for the $C^{(0)}BF$ component. The couplings in eq.(3.33) of \cite{Becker:2011ar} in the momentum space and for $D_3$-brane in string frame is 
\beqa
{\cal L}^{(4)}_{CBA}&=&i\frac{\pi^3}{6}C\eps^{a_1a_2a_3a_4}\bigg[-4i(p_2.k_3)^2B_{a_1a_2}F_{a_3a_4}-2p_2.k_3\,p_2.V.H_{a_1a_2}F_{a_1a_2}\nonumber\\
&&+2p_2.k_3\,H_{a_1a_2b}k_3^bF_{a_3a_4}+p_2.V.p_2\,H_{a_1a_2b}k_3^bF_{a_3a_4}\nonumber\\
&&-\frac{2}{3}p_2.V.p_2\,H_{a_1a_2a_3}p_2^bF_{ba_4}-\frac{4}{3}p_2.k_3\,H_{a_1a_2a_3}p_2^bF_{ba_4}\nonumber\\
&&+2p_1.N.p_2\,p_2.V.H_{a_1a_2}F_{a_3a_4}-p_2.V.p_2\,p_1.N.H_{a_1a_2}F_{a_3a_4}\bigg]\labell{L4}
\eeqa
where we have used  our convention that $\alpha'=2$ and the gauge invariant combination of $B$ and $F$ is $B+2\pi\alpha' F$. The terms in the second line are zero using the relation \reef{iden.2}. Using the identity \reef{iden.3}, one can rewrite the terms in the third line  as
\beqa
-2p_2.V.p_2\, p_2.V.H_{a_1a_2}(*F)^{a_1a_2}-4p_2.k_3\,p_2.V.H_{a_1a_2}(*F)^{a_1a_2}\nonumber
\eeqa
Then it is easy to verify that the couplings \reef{L4} reduces to the corresponding couplings in \reef{final}. This fixes the constant factor to be  $\alpha=-i/4$.

{\bf Acknowledgments}:   This work is supported by Ferdowsi University of Mashhad under grant 2/200085-1390/09/27.

\newpage
{\bf\Large {Appendix: Some Integrals}}

In this appendix, we  write the integrals that appear in the explicit calculation of the amplitudes \reef{amp21} and \reef{amp24}. The following integrals appear in amplitude \reef{amp21}:
\beqa
I_{1}&=&\frac{K}{z_{12} z_{31} z_{3 \bar{1}} z_{2 \bar{2}} z_{\bar{1} \bar{2}}}\labell{apen}\\
I_3&=&\frac{K}{z_{12} z_{32} z_{1\bar{1}} z_{3 \bar{1}} z_{1\bar{2}}}-\frac{2 K}{z_{12} z_{31} z_{2 \bar{1}} z_{3 \bar{1}} z_{2 \bar{2}}}+\frac{K}{z_{12} z_{31} z_{2 \bar{1}} z_{3 \bar{1}} z_{3 \bar{2}}}+\frac{K}{z_{31} z_{32} z_{1\bar{1}} z_{2 \bar{1}} z_{\bar{1} \bar{2}}}\nonumber\\
I_4&=&\frac{x_3^2 \left(z_{1\bar{1}}+z_{2 \bar{2}}\right)-z_2 \bar{z}_1 \bar{z}_2+2 x_3 \left(-z_1 z_2+\bar{z}_1 \bar{z}_2\right)+z_1 \left(-\bar{z}_1 \bar{z}_2+z_2 \left(\bar{z}_1+\bar{z}_2\right)\right)}{z_{12} z_{31} z_{32} z_{2 \bar{1}} z_{3 \bar{1}} z_{1\bar{2}} z_{3 \bar{2}} z_{\bar{1} \bar{2}}}K\nonumber\\
I_5&=&\frac{K}{z_{12} z_{31} z_{32} z_{1\bar{1}} z_{2 \bar{1}} z_{3 \bar{1}} z_{1\bar{2}} z_{3 \bar{2}} z_{\bar{1} \bar{2}}}\left(-x_3^2 z_1^2+x_3^2 z_1 z_2+z_1^2 z_2 z_{\bar{1} \bar{2}}+2 x_3 z_1^2 \bar{z}_1+x_3^2 z_2 \bar{z}_1\right.\nonumber\\
&&\left.-4 x_3 z_1 z_2 \bar{z}_1-x_3^2 \bar{z}_1^2+2 x_3 z_1 \bar{z}_1^2-2 z_1^2 \bar{z}_1^2+z_1 z_2 \bar{z}_1^2+x_3^2 z_1 \bar{z}_2-2 x_3^2 z_2 \bar{z}_2+2 x_3 z_1 z_2 \bar{z}_2\right.\nonumber\\
&&\left.+x_3^2 \bar{z}_1 \bar{z}_2-4 x_3 z_1 \bar{z}_1 \bar{z}_2+z_1^2 \bar{z}_1 \bar{z}_2+2 x_3 z_2 \bar{z}_1 \bar{z}_2+z_1 \bar{z}_1^2 \bar{z}_2-z_2 \bar{z}_1^2 \bar{z}_2\right)\nonumber\\
I_6&=&\frac{-z_2 \bar{z}_1 \bar{z}_2+x_3^2 \left(z_{12}-z_{\bar{1}\bar{2}}\right)+2 x_3 \left(z_2 \bar{z}_1-z_1 \bar{z}_2\right)+z_1 \left(-z_2 \bar{z}_1+\left(z_2+\bar{z}_1\right) \bar{z}_2\right)}{z_{31} z_{32} z_{1\bar{1}} z_{2 \bar{1}} z_{3 \bar{1}} z_{1\bar{2}} z_{2 \bar{2}} z_{3 \bar{2}}}K\nonumber\\
I_7&=&\frac{-z_2 \bar{z}_1 \bar{z}_2+x_3^2 \left(z_{12}-z_{\bar{1}\bar{2}}\right)+2 x_3 \left(z_2 \bar{z}_1-z_1 \bar{z}_2\right)+z_1 \left(-z_2 \bar{z}_1+\left(z_2+\bar{z}_1\right) \bar{z}_2\right)}{z_{12} z_{31} z_{32} z_{1\bar{1}} z_{3 \bar{1}} z_{2 \bar{2}} z_{3 \bar{2}} z_{\bar{1} \bar{2}}}K\nonumber\\
I_8&=&\frac{z_{2 \bar{2}} }{z_{12} z_{32} z_{2 \bar{1}} z_{1\bar{2}} z_{3 \bar{2}} z_{\bar{1} \bar{2}}}K\nonumber\\
I_9&=&\frac{x_3^2 \left(z_{1\bar{1}}+z_{2 \bar{2}}\right)-z_2 \bar{z}_1 \bar{z}_2+2 x_3 \left(-z_1 z_2+\bar{z}_1 \bar{z}_2\right)+z_1 \left(-\bar{z}_1 \bar{z}_2+z_2 \left(\bar{z}_1+\bar{z}_2\right)\right)}{z_{31} z_{32} z_{1\bar{1}}z_{2 \bar{1}} z_{3 \bar{1}} z_{1\bar{2}} z_{2 \bar{2}} z_{3 \bar{2}}}K\nonumber\\
I_{10}&=&\frac{-z_2 \bar{z}_1 \bar{z}_2+x_3^2 \left(z_{12}-z_{\bar{1} \bar{2}}\right)+2 x_3 \left(z_2 \bar{z}_1-z_1 \bar{z}_2\right)+z_1 \left(-z_2 \bar{z}_1+\left(z_2+\bar{z}_1\right) \bar{z}_2\right)}{z_{12} z_{31} z_{32} z_{2 \bar{1}} z_{3 \bar{1}} z_{1\bar{2}} z_{3 \bar{2}} z_{\bar{1} \bar{2}}}K\nonumber
\eeqa
Using these integrals, one can verify the relations \reef{rel.1} which have been found by imposing the Ward identity on the amplitude \reef{amp20}. The above integrals, however, do not appear in the final form of the amplitude \reef{ampg}.

The integrals in the R-R amplitude \reef{amp24} are
\beqa
 J_1&=&\frac{z_1 z_2-4 z_1 \bar{z}_1+3 z_2 \bar{z}_1+\left(3 z_1-4 z_2+\bar{z}_1\right) \bar{z}_2}{z_{12} z_{31} z_{2 \bar{1}} z_{3 \bar{1}} z_{1\bar{2}} z_{2 \bar{2}} z_{\bar{1} \bar{2}}}K\nonumber\\
 J_2&=&\frac{\left(x_3 \left(z_{12}+z_{1\bar{2}}\right)+2 z_2 \bar{z}_2-z_1 \left(z_2+\bar{z}_2\right)\right) \left(z_2 \left(\bar{z}_1-2 \bar{z}_2\right)+\bar{z}_1 \bar{z}_2+x_3 \left(z_{2\bar{1}}+z_{\bar{2} \bar{1}}\right)\right)}{z_{12} z_{31} z_{32} z_{2 \bar{1}} z_{3 \bar{1}} z_{1\bar{2}} z_{2 \bar{2}} z_{3 \bar{2}} z_{\bar{1} \bar{2}}}K\nonumber\\
 J_3&=&\frac{z_2 \left(\bar{z}_1-4 \bar{z}_2\right)+3 \bar{z}_1 \bar{z}_2+z_1 \left(3 z_2-4 \bar{z}_1+\bar{z}_2\right)}{z_{12} z_{31} z_{2 \bar{1}} z_{3 \bar{1}}z_{1\bar{2}} z_{2 \bar{2}} z_{\bar{1} \bar{2}}}K\nonumber\\
 J_4&=&\frac{z_2 \left(\bar{z}_1-2 \bar{z}_2\right)+\bar{z}_1 \bar{z}_2+z_1 \left(z_{2\bar{1}}+z_{\bar{2} \bar{1}}\right)}{z_{12} z_{31} z_{2 \bar{1}} z_{3 \bar{1}} z_{1\bar{2}} z_{2 \bar{2}} z_{\bar{1} \bar{2}}}K
 \eeqa
Using the above integrals, one can verify the relations in \reef{rel.3}. The above integrals, however, do not appear in the final form of the amplitude \reef{amp34}. The following integrals appear in the amplitudes in section 4.1:
\beqa
I&=& \frac{\frac{1}{z_{32}^2 z_{1 \bar{2}} z_{\bar{1} \bar{2}}}-\frac{1}{z_{12} z_{2 \bar{1}} z_{3 \bar{2}}^2}}{z_{1 \bar{1}}} K\nonumber\\
J_0&=&\frac{z_1 \left(z_2-2 \bar{z}_1\right)+z_2 \bar{z}_1+x_3 \left(z_1-2 z_2+\bar{z}_1\right)}{z_{12} z_{31} z_{32} z_{2 \bar{1}} z_{3 \bar{1}} z_{1 \bar{2}} z_{\bar{1} \bar{2}}}K\nonumber\\
J_5&=&\frac{z_{3 \bar{2}}}{z_{31} z_{32} z_{3 \bar{1}} z_{2 \bar{2}} z_{1 \bar{2}} z_{\bar{1} \bar{2}}}K\nonumber\\
J_6&=&\frac{\left(x_3^2 \left(z_{12}+\bar{z}_1-\bar{z}_2\right)-z_2 \bar{z}_1 \bar{z}_2+2 x_3 \left(-z_1 \bar{z}_1+z_2 \bar{z}_2\right)+z_1 \left(z_2 z_{\bar{1} \bar{2}}+\bar{z}_1 \bar{z}_2\right)\right){}^2}{z_{12} z_{31} z_{32}^2 z_{2 \bar{1}} z_{3 \bar{1}} z_{1 \bar{1}} z_{3 \bar{2}}^2 z_{1 \bar{2}} z_{\bar{1} \bar{2}}}K\nonumber\\
J_7&=&\left(-z_2 \bar{z}_1 \bar{z}_2 \left(z_2 \left(\bar{z}_1-2 \bar{z}_2\right)+\bar{z}_1 \bar{z}_2\right)+2 z_1 \bar{z}_2 \left(z_2^2 \bar{z}_1-\left(z_2^2+z_2 \bar{z}_1-\bar{z}_1^2\right) \bar{z}_2\right)\right.\nonumber\\
&&\left.+x_3 \left(z_2 \left(z_1^2 z_2+2 z_1 z_2 \bar{z}_1+\left(-4 z_1+z_2\right) \bar{z}_1^2\right)-4 z_{2 \bar{1}} \left(z_1^2+z_2 \bar{z}_1\right) \bar{z}_2\right.\right.\nonumber\\
&&\left.\left.-\left(z_1^2+\bar{z}_1^2+2 z_1 \left(-2 z_2+\bar{z}_1\right)\right) \bar{z}_2^2\right)+x_3^2 \left(-z_2 \bar{z}_1^2+2 z_1 \left(-z_2^2+z_2 \bar{z}_1+z_{\bar{1} \bar{2}} \bar{z}_1\right)\right.\right.\nonumber\\
&&\left.\left.+\left(2 z_2^2-\bar{z}_1^2\right) \bar{z}_2+2 \left(-z_2+\bar{z}_1\right) \bar{z}_2^2+z_1^2 \left(z_2-2 \bar{z}_1+\bar{z}_2\right)\right)\right.\nonumber\\
&&\left.+z_1^2 \left(-2 \bar{z}_1^2 \bar{z}_2+z_2^2 \left(-2 \bar{z}_1+\bar{z}_2\right)+z_2 \left(2 \bar{z}_1^2+\bar{z}_2^2\right)\right)\right)K/\left(z_{12} z_{31} z_{32} z_{2 \bar{1}} z_{3 \bar{1}} z_{1 \bar{1}} z_{2 \bar{2}} z_{3 \bar{2}} z_{1 \bar{2}} z_{\bar{1} \bar{2}}\right)\nonumber\\
J_8&=&\left(-z_2 \bar{z}_1 \bar{z}_2 \left(z_2 \left(\bar{z}_1-2 \bar{z}_2\right)+\bar{z}_1 \bar{z}_2\right)+z_1^2 \left(-2 z_2 \bar{z}_1^2+\left(z_2^2+2 \bar{z}_1^2\right) \bar{z}_2+\left(z_2-2 \bar{z}_1\right) \bar{z}_2^2\right)\right.\nonumber\\
&&\left.+2 z_1 z_2 \left(\bar{z}_1 \bar{z}_2^2+z_2 \left(\bar{z}_1^2-\bar{z}_1 \bar{z}_2-\bar{z}_2^2\right)\right)+x_3^2 \left(z_2 \left(2 z_2-\bar{z}_1\right) \bar{z}_1-\left(2 z_2^2+\bar{z}_1^2\right) \bar{z}_2+2 z_2 \bar{z}_2^2\right.\right.\nonumber\\
&&\left.\left.+z_1^2 \left(z_2-2 \bar{z}_1+\bar{z}_2\right)+2 z_1 \left(-z_2 \bar{z}_1+\bar{z}_1^2+\bar{z}_1 \bar{z}_2-\bar{z}_2^2\right)\right)+x_3 \left(z_1^2 \left(-z_2^2+4 z_2 z_{\bar{1} \bar{2}}+\bar{z}_2^2\right)\right.\right.\nonumber\\
&&\left.\left.+\bar{z}_1 \left(-z_2^2 \bar{z}_1+4 z_2 \bar{z}_1 \bar{z}_2+\left(-4 z_2+\bar{z}_1\right) \bar{z}_2^2\right)+2 z_1 \left(-z_2^2 \left(\bar{z}_1-2 \bar{z}_2\right)+\bar{z}_1 \bar{z}_2 \left(-2 \bar{z}_1+\bar{z}_2\right)\right)\right)\right)K\nonumber\\
&&/\left(z_{12} z_{31} z_{32} z_{2 \bar{1}} z_{3 \bar{1}} z_{1 \bar{1}} z_{2 \bar{2}} z_{3 \bar{2}} z_{1 \bar{2}} z_{\bar{1} \bar{2}}\right) \nonumber\\
J_9&=&\frac{z_2 \left(\bar{z}_1-2 \bar{z}_2\right)+\bar{z}_1 \bar{z}_2+z_1 \left(z_2-2 \bar{z}_1+\bar{z}_2\right)}{z_{12} z_{32} z_{2 \bar{1}} z_{1 \bar{1}} z_{3 \bar{2}} z_{1 \bar{2}} z_{\bar{1} \bar{2}}}K\nonumber\\
J_{10}&=&\frac{z_2 \left(\bar{z}_1-4 \bar{z}_2\right)+3 \bar{z}_1 \bar{z}_2+z_1 \left(3 z_2-4 \bar{z}_1+\bar{z}_2\right)}{z_{12} z_{31} z_{2 \bar{1}} z_{3 \bar{1}} z_{2 \bar{2}} z_{1 \bar{2}} z_{\bar{1} \bar{2}}}K\nonumber\\
J_{11}&=&\frac{z_2 \left(\bar{z}_1-4 \bar{z}_2\right)+3 \bar{z}_1 \bar{z}_2+z_1 \left(3 z_2-4 \bar{z}_1+\bar{z}_2\right)}{z_{12} z_{32} z_{2 \bar{1}} z_{1 \bar{1}} z_{3 \bar{2}} z_{1 \bar{2}} z_{\bar{1} \bar{2}}}K\nonumber\\
J_{12}&=&\frac{z_1 z_2-4 z_1 \bar{z}_1+3 z_2 \bar{z}_1+\left(3 z_1-4 z_2+\bar{z}_1\right) \bar{z}_2}{z_{12} z_{31} z_{2 \bar{1}} z_{3 \bar{1}} z_{2 \bar{2}} z_{1 \bar{2}} z_{\bar{1} \bar{2}}}K\nonumber\\
J_{13}&=&\frac{z_1 z_2-4 z_1 \bar{z}_1+3 z_2 \bar{z}_1+\left(3 z_1-4 z_2+\bar{z}_1\right) \bar{z}_2}{z_{12} z_{32} z_{2 \bar{1}} z_{1 \bar{1}} z_{3 \bar{2}} z_{1 \bar{2}} z_{\bar{1} \bar{2}}}K\nonumber\\
J_{14}&=&\frac{z_{31} z_{3 \bar{1}} z_{2 \bar{2}}^2}{z_{12} z_{32}^2 z_{2 \bar{1}} z_{1 \bar{1}} z_{3 \bar{2}}^2 z_{1 \bar{2}} z_{\bar{1} \bar{2}}}K\nonumber\\
J_{15}&=&J\,=\,\frac{z_{1 \bar{1}}}{z_{12} z_{31} z_{2 \bar{1}} z_{3 \bar{1}} z_{1 \bar{2}} z_{\bar{1} \bar{2}}}K\nonumber\\
J_{17}&=&-J_{16}\,=\,\frac{z_{2 \bar{2}}}{z_{12} z_{32} z_{2 \bar{1}} z_{3 \bar{2}} z_{1 \bar{2}} z_{\bar{1} \bar{2}}}K\nonumber\\
J_{18}&=&\left(\left(z_1 \left(z_2-2 \bar{z}_1\right)+z_2 \bar{z}_1+x_3 \left(z_{12} -z_{2\bar{1}} \right)\right) \left(-2 z_1 \bar{z}_1+x_3 \left(z_1+\bar{z}_1-2 \bar{z}_2\right)+\left(z_1+\bar{z}_1\right) \bar{z}_2\right)\right)K\nonumber\\
&&/\left(z_{12} z_{31} z_{32} z_{2 \bar{1}} z_{3 \bar{1}} z_{1 \bar{1}} z_{3 \bar{2}} z_{1 \bar{2}} z_{\bar{1} \bar{2}}\right)\nonumber\\
J_{19}&=&\frac{\left(z_2 \left(\bar{z}_1-2 \bar{z}_2\right)+\bar{z}_1 \bar{z}_2+z_1 \left(z_2-2 \bar{z}_1+\bar{z}_2\right)\right){}^2}{z_{12} z_{31} z_{2 \bar{1}} z_{3 \bar{1}} z_{1 \bar{1}} z_{2 \bar{2}}^2 z_{1 \bar{2}} z_{\bar{1} \bar{2}}}K\nonumber\\
J_{20}&=&\frac{K}{z_{31} z_{3 \bar{1}} z_{1 \bar{1}} z_{2 \bar{2}}^2}
\eeqa
Using the above integrals, one can find the relations between the integrals that appear in section 4.1.

%%%%%%%%%%%%%%%%%%%%%%%%%%%%%%%%%%%%%%%%

%%%%%%%%%%%%%%%%%%%%%%%%%%%%%%%%%%%%%%%
%\beqa -\frac{\hbar^2}{2\mu}\frac{\partial^2}{\partial x^2} \eeqa
%\newpage


\begin{thebibliography}{99}


%\cite{Font:1990gx}
\bibitem{Font:1990gx}
  A.~Font, L.~E.~Ibanez, D.~Lust and F.~Quevedo,
  %``Strong - weak coupling duality and nonperturbative effects in string
  %theory,''
  Phys.\ Lett.\  B {\bf 249}, 35 (1990).
  %%CITATION = PHLTA,B249,35;%%
  %\cite{Rey:1989xj}
\bibitem{Rey:1989xj}
  S.~J.~Rey,
  %``THE CONFINING PHASE OF SUPERSTRINGS AND AXIONIC STRINGS,''
  Phys.\ Rev.\  D {\bf 43}, 526 (1991).
  %%CITATION = PHRVA,D43,526;%%

%\cite{Sen:1994fa}
\bibitem{Sen:1994fa}
  A.~Sen,
  %``Strong - weak coupling duality in four-dimensional string theory,''
  Int.\ J.\ Mod.\ Phys.\  A {\bf 9}, 3707 (1994)
  [arXiv:hep-th/9402002].
  %%CITATION = IMPAE,A9,3707;%%
%\cite{Sen:1994yi}
\bibitem{Sen:1994yi}
  A.~Sen,
  %``Dyon - monopole bound states, selfdual harmonic forms on the multi -
  %monopole moduli space, and SL(2,Z) invariance in string theory,''
  Phys.\ Lett.\  B {\bf 329}, 217 (1994)
  [arXiv:hep-th/9402032].
  %%CITATION = PHLTA,B329,217;%%

%\cite{Schwarz:1993cr}
\bibitem{Schwarz:1993cr}
  J.~H.~Schwarz,
  %``Does string theory have a duality symmetry relating weak and strong
  %coupling?,''
  arXiv:hep-th/9307121.
  %%CITATION = HEP-TH/9307121;%%

%\cite{Hull:1994ys}
\bibitem{Hull:1994ys}
  C.~M.~Hull and P.~K.~Townsend,
  %``Unity of superstring dualities,''
  Nucl.\ Phys.\  B {\bf 438}, 109 (1995)
  [arXiv:hep-th/9410167].
  %%CITATION = NUPHA,B438,109;%%
 %\cite{Becker:2007zj}
\bibitem{Becker:2007zj}
  K.~Becker, M.~Becker and J.~H.~Schwarz,
  ``String theory and M-theory: A modern introduction,''
%\href{http://www.slac.stanford.edu/spires/find/hep/www?irn=7073143}{SPIRES entry}
{\it  Cambridge, UK: Cambridge Univ. Pr. (2007) 739 p} 
 
 
%\cite{Garousi:2011we}
\bibitem{Garousi:2011we} 
  M.~R.~Garousi,
  %``S-duality of S-matrix,''  
  JHEP {\bf 1111}, 016 (2011)  [arXiv:1106.1714 [hep-th]].  
  %%CITATION = ARXIV:1106.1714;%%
 
% ******** 
  
%\cite{Green:1997tv}
\bibitem{Green:1997tv}
  M.~B.~Green and M.~Gutperle,
  %``Effects of D instantons,''
  Nucl.\ Phys.\  B {\bf 498}, 195 (1997)
  [arXiv:hep-th/9701093].
  %%CITATION = NUPHA,B498,195;%%
%\cite{Green:1997di}
\bibitem{Green:1997di}
  M.~B.~Green and P.~Vanhove,
  %``D instantons, strings and M theory,''
  Phys.\ Lett.\  B {\bf 408}, 122 (1997)
  [arXiv:hep-th/9704145].
  %%CITATION = PHLTA,B408,122;%%
%\cite{Green:1997as}
\bibitem{Green:1997as}
  M.~B.~Green, M.~Gutperle and P.~Vanhove,
  %``One loop in eleven-dimensions,''
  Phys.\ Lett.\  B {\bf 409}, 177 (1997)
  [arXiv:hep-th/9706175].
  %%CITATION = PHLTA,B409,177;%%
%\cite{Kiritsis:1997em}
\bibitem{Kiritsis:1997em}
  E.~Kiritsis and B.~Pioline,
  %``On R**4 threshold corrections in IIb string theory and (p, q) string
  %instantons,''
  Nucl.\ Phys.\  B {\bf 508}, 509 (1997)
  [arXiv:hep-th/9707018].
  %%CITATION = NUPHA,B508,509;%%
%\cite{Green:1997me}
\bibitem{Green:1997me}
  M.~B.~Green, M.~Gutperle and H.~h.~Kwon,
  %``Sixteen fermion and related terms in M theory on T**2,''
  Phys.\ Lett.\  B {\bf 421}, 149 (1998)
  [arXiv:hep-th/9710151].
  %%CITATION = PHLTA,B421,149;%%
%\cite{Pioline:1998mn}
\bibitem{Pioline:1998mn}
  B.~Pioline,
  %``A Note on nonperturbative R**4 couplings,''
  Phys.\ Lett.\  B {\bf 431}, 73 (1998)
  [arXiv:hep-th/9804023].
  %%CITATION = PHLTA,B431,73;%%
%\cite{Green:1998by}
\bibitem{Green:1998by}
  M.~B.~Green and S.~Sethi,
  %``Supersymmetry constraints on type IIB supergravity,''
  Phys.\ Rev.\  D {\bf 59}, 046006 (1999)
  [arXiv:hep-th/9808061].
  %%CITATION = PHRVA,D59,046006;%%
%\cite{Green:1999pu}
\bibitem{Green:1999pu}
  M.~B.~Green, H.~h.~Kwon and P.~Vanhove,
  %``Two loops in eleven-dimensions,''
  Phys.\ Rev.\  D {\bf 61}, 104010 (2000)
  [arXiv:hep-th/9910055].
  %%CITATION = PHRVA,D61,104010;%%
%\cite{Obers:1999es}
\bibitem{Obers:1999es}
  N.~A.~Obers and B.~Pioline,
  %``Eisenstein series in string theory,''
  Class.\ Quant.\ Grav.\  {\bf 17}, 1215 (2000)
  [arXiv:hep-th/9910115].
  %%CITATION = CQGRD,17,1215;%%

%\cite{Sinha:2002zr}
\bibitem{Sinha:2002zr}
  A.~Sinha,
  %``The G(hat)**4 lambda**16 term in IIB supergravity,''
  JHEP {\bf 0208}, 017 (2002)
  [arXiv:hep-th/0207070].
  %%CITATION = JHEPA,0208,017;%%
%\cite{Berkovits:2004px}
\bibitem{Berkovits:2004px}
  N.~Berkovits,
  %``Multiloop amplitudes and vanishing theorems using the pure spinor formalism
  %for the superstring,''
  JHEP {\bf 0409}, 047 (2004)
  [arXiv:hep-th/0406055].
  %%CITATION = JHEPA,0409,047;%%
%\cite{D'Hoker:2005jc}
\bibitem{D'Hoker:2005jc}
  E.~D'Hoker and D.~H.~Phong,
  %``Two-loop superstrings VI: Non-renormalization theorems and the 4-point
  %function,''
  Nucl.\ Phys.\  B {\bf 715}, 3 (2005)
  [arXiv:hep-th/0501197].
  %%CITATION = NUPHA,B715,3;%%
%\cite{D'Hoker:2005ht}
\bibitem{D'Hoker:2005ht}
  E.~D'Hoker, M.~Gutperle and D.~H.~Phong,
  %``Two-loop superstrings and S-duality,''
  Nucl.\ Phys.\  B {\bf 722}, 81 (2005)
  [arXiv:hep-th/0503180].
  %%CITATION = NUPHA,B722,81;%%
%\cite{Green:2005ba}
\bibitem{Green:2005ba}
  M.~B.~Green and P.~Vanhove,
  %``Duality and higher derivative terms in M theory,''
  JHEP {\bf 0601}, 093 (2006)
  [arXiv:hep-th/0510027].
  %%CITATION = JHEPA,0601,093;%%
%\cite{Green:2006gt}
\bibitem{Green:2006gt}
  M.~B.~Green, J.~G.~Russo and P.~Vanhove,
  %``Non-renormalisation conditions in type II string theory and maximal
  %supergravity,''
  JHEP {\bf 0702}, 099 (2007)
  [arXiv:hep-th/0610299].
  %%CITATION = JHEPA,0702,099;%%
%\cite{Basu:2007ru}
\bibitem{Basu:2007ru}
  A.~Basu,
  %``The D**4 R**4 term in type IIB string theory on T**2 and U-duality,''
  Phys.\ Rev.\  D {\bf 77}, 106003 (2008)
  [arXiv:0708.2950 [hep-th]].
  %%CITATION = PHRVA,D77,106003;%%
%\cite{Basu:2007ck}
\bibitem{Basu:2007ck}
  A.~Basu,
  %``The D**6 R**4 term in type IIB string theory on T**2 and U-duality,''
  Phys.\ Rev.\  D {\bf 77}, 106004 (2008)
  [arXiv:0712.1252 [hep-th]].
  %%CITATION = PHRVA,D77,106004;%%




%***************

 %\cite{Bachas:1999um}
\bibitem{Bachas:1999um}
  C.~P.~Bachas, P.~Bain and M.~B.~Green,
  %``Curvature terms in D-brane actions and their M-theory origin,''
  JHEP {\bf 9905}, 011 (1999)
  [arXiv:hep-th/9903210].
  %%CITATION = JHEPA,9905,011;%%
%\cite{Basu:2008gt}
\bibitem{Basu:2008gt} 
  A.~Basu,
  %``Constraining the D3-brane effective action,''
  JHEP {\bf 0809}, 124 (2008)
  [arXiv:0808.2060 [hep-th]].
  %%CITATION = JHEPA,0809,124;%%

%\cite{Garousi:2011fc}
\bibitem{Garousi:2011fc}
  M.~R.~Garousi,
  %``S-duality of D-brane action at order $O(\alpha'^2)$,''
  Phys.\ Lett.\  B {\bf 701}, 465 (2011)
  [arXiv:1103.3121 [hep-th]].
  %%CITATION = PHLTA,B701,465;%%

%\cite{Garousi:2011vs}
\bibitem{Garousi:2011vs} 
  M.~R.~Garousi,
  %``On S-duality of D$_3$-brane S-matrix,'' 
  Phys.\ Rev.\ D {\bf 84}, 126019 (2011) 
  [arXiv:1108.4782 [hep-th]].  
  %%CITATION = ARXIV:1108.4782;%%


%\cite{Garousi:2011jh}
\bibitem{Garousi:2011jh} 
  M.~R.~Garousi,
  %``S-duality of non-abelian S-matrix,''  
  arXiv:1109.5555 [hep-th].  
  %%CITATION = ARXIV:1109.5555;%%
  %\cite{Cohn:1986bn}
\bibitem{Cohn:1986bn} 
  J.~Cohn, D.~Friedan, Z.~-a.~Qiu and S.~H.~Shenker,
  %``Covariant Quantization Of Supersymmetric String Theories: The Spinor Field Of The Ramond-neveu-schwarz Model,''  
  Nucl.\ Phys.\ B {\bf 278}, 577 (1986).  
  %%CITATION = NUPHA,B278,577;%%

  
  %\cite{Fotopoulos:2001pt}
\bibitem{Fotopoulos:2001pt}
  A.~Fotopoulos,
  %``On (alpha-prime)**2 corrections to the D-brane action for nongeodesic world
  %volume embeddings,''
  JHEP {\bf 0109}, 005 (2001)
  [arXiv:hep-th/0104146].
  %%CITATION = JHEPA,0109,005;%%
  
%\cite{Garousi:1996ad}
\bibitem{Garousi:1996ad}
  M.~R.~Garousi and R.~C.~Myers,
  %``Superstring Scattering from D-Branes,''
  Nucl.\ Phys.\  B {\bf 475}, 193 (1996)
  [arXiv:hep-th/9603194].
  %%CITATION = NUPHA,B475,193;%%
 %\cite{Hashimoto:1996bf}
\bibitem{Hashimoto:1996bf}
  A.~Hashimoto and I.~R.~Klebanov,
  %``Scattering of strings from D-branes,''
  Nucl.\ Phys.\ Proc.\ Suppl.\  {\bf 55B} (1997) 118
  [arXiv:hep-th/9611214].
  %%CITATION = NUPHZ,55B,118;%% 
  %\cite{Stieberger:2009hq}
\bibitem{Stieberger:2009hq} 
  S.~Stieberger,
  %``Open & Closed vs. Pure Open String Disk Amplitudes,''  
  arXiv:0907.2211 [hep-th].  
  %%CITATION = ARXIV:0907.2211;%%
%\cite{Kitazawa:1987xj}
\bibitem{Kitazawa:1987xj} 
  Y.~Kitazawa,
  %``Effective Lagrangian For Open Superstring From Five Point Function,''  
  Nucl.\ Phys.\ B {\bf 289}, 599 (1987).  
  %%CITATION = NUPHA,B289,599;%%

%\cite{ISG}
\bibitem{ISG}
I.S. Gradshteyn and I.M. Ryzhik, Table of Integrals, Series and Products, Academic Press 1994.

 
 %\cite{Huber:2005yg}
\bibitem{Huber:2005yg}
  T.~Huber and D.~Maitre,
  %``HypExp, a Mathematica package for expanding hypergeometric functions
  %around integer-valued parameters,''
  Comput.\ Phys.\ Commun.\  {\bf 175}, 122 (2006)
  [arXiv:hep-ph/0507094].
  %%CITATION = CPHCB,175,122;%%

%************  
  
 %\cite{Billo:1998vr}
\bibitem{Billo:1998vr}
  M.~Billo, P.~Di Vecchia, M.~Frau, A.~Lerda, I.~Pesando, R.~Russo and S.~Sciuto,
  %``Microscopic string analysis of the D0-D8 brane system and dual R-R
  %states,''
  Nucl.\ Phys.\  B {\bf 526}, 199 (1998)
  [arXiv:hep-th/9802088].
  %%CITATION = NUPHA,B526,199;%%  
  
  


 %\cite{Liu:2001qa}
\bibitem{Liu:2001qa}
  H.~Liu and J.~Michelson,
  %``*-trek III: The search for Ramond-Ramond couplings,''
  Nucl.\ Phys.\  B {\bf 614}, 330 (2001)
  [arXiv:hep-th/0107172].
  %%CITATION = NUPHA,B614,330;%%
%\cite{Garousi:2008ge}
\bibitem{Garousi:2008ge}
  M.~R.~Garousi and E.~Hatefi,
  %``More on WZ action of non-BPS branes,''
  JHEP {\bf 0903}, 008 (2009)
  [arXiv:0812.4216 [hep-th]].
  %%CITATION = JHEPA,0903,008;%%
 %\cite{Garousi:2010bm}
\bibitem{Garousi:2010bm} 
  M.~R.~Garousi and M.~Mir,
  %``On RR couplings on D-branes at order $O(\alpha'^2)$,''  
  JHEP {\bf 1102}, 008 (2011)  [arXiv:1012.2747 [hep-th]].  
  %%CITATION = ARXIV:1012.2747;%%
 
 %\cite{Gibbons:1995ap}
\bibitem{Gibbons:1995ap}
  G.~W.~Gibbons and D.~A.~Rasheed,
  %``Sl(2,R) invariance of nonlinear electrodynamics coupled to an axion and a
  %dilaton,''
  Phys.\ Lett.\  B {\bf 365}, 46 (1996)
  [arXiv:hep-th/9509141].
  %%CITATION = PHLTA,B365,46;%%
%\cite{Tseytlin:1996it}
\bibitem{Tseytlin:1996it}
  A.~A.~Tseytlin,
  %``Selfduality of Born-Infeld action and Dirichlet three-brane of type IIB
  %superstring theory,''
  Nucl.\ Phys.\  B {\bf 469}, 51 (1996)
  [arXiv:hep-th/9602064].
  %%CITATION = NUPHA,B469,51;%%
%\cite{Green:1996qg}
\bibitem{Green:1996qg}
  M.~B.~Green and M.~Gutperle,
  %``Comments on three-branes,''
  Phys.\ Lett.\  B {\bf 377}, 28 (1996)
  [arXiv:hep-th/9602077].
  %%CITATION = PHLTA,B377,28;%%  
 %\cite{Hartl:2009yf}
\bibitem{Hartl:2009yf}
  D.~Hartl, O.~Schlotterer and S.~Stieberger,
  %``Higher Point Spin Field Correlators in D=4 Superstring Theory,''
  Nucl.\ Phys.\  B {\bf 834}, 163 (2010)
  [arXiv:0911.5168 [hep-th]].
  %%CITATION = NUPHA,B834,163;%%

%\cite{Hartl:2010ks}
\bibitem{Hartl:2010ks} 
  D.~Haertl and O.~Schlotterer,
  %``Higher Loop Spin Field Correlators in Various Dimensions,''  
  Nucl.\ Phys.\ B {\bf 849}, 364 (2011)  [arXiv:1011.1249 [hep-th]].  
  %%CITATION = ARXIV:1011.1249;%%
  
  
%\cite{Becker:2011ar}
\bibitem{Becker:2011ar}
  K.~Becker, G.~Guo and D.~Robbins,
  %``Four-Derivative Brane Couplings from String Amplitudes,''
  arXiv:1110.3831 [hep-th].
  %%CITATION = ARXIV:1110.3831;%%

 %\cite{Hashimoto:1996kf}
\bibitem{Hashimoto:1996kf}
  A.~Hashimoto and I.~R.~Klebanov,
  %``Decay of excited D-branes,''
  Phys.\ Lett.\  B {\bf 381}, 437 (1996)
  [arXiv:hep-th/9604065].
  %%CITATION = PHLTA,B381,437;%%


 %\cite{Garousi:1998fg}
\bibitem{Garousi:1998fg}
  M.~R.~Garousi and R.~C.~Myers,
  %``World volume interactions on D-branes,''
  Nucl.\ Phys.\  B {\bf 542}, 73 (1999)
  [arXiv:hep-th/9809100].
  %%CITATION = NUPHA,B542,73;%%
  
%\cite{Garousi:2011ut}
\bibitem{Garousi:2011ut}
  M.~R.~Garousi and M.~Mir,
  %``Towards extending the Chern-Simons couplings at order $O(\alpha'^2)$,''
  JHEP {\bf 1105}, 066 (2011)
  [arXiv:1102.5510 [hep-th]].
  %%CITATION = JHEPA,1105,066;%%
 

  
\end{thebibliography}
\end{document}